\documentclass[12pt]{article}
\usepackage{epsfig, epsf, graphicx, subfigure}
\usepackage{pstricks, pst-node, psfrag}
\usepackage{amssymb,amsmath}
\usepackage{verbatim,enumerate}
\usepackage{rotating, lscape}
\usepackage{setspace}
\usepackage{multicol}
\usepackage{multirow}
\usepackage{gensymb} 
\usepackage{natbib}
\usepackage{mathtools}
\usepackage{booktabs}												
\usepackage[utf8]{inputenc}
\usepackage{hyperref}
\usepackage[]{algorithm2e}
\usepackage[toc,page]{appendix}
\usepackage{amssymb}
\usepackage{pifont}
\def\log{\hbox{log}}

\def\boxit#1{\vbox{\hrule\hbox{\vrule\kern6pt
          \vbox{\kern6pt#1\kern6pt}\kern6pt\vrule}\hrule}}

\def\bse{\begin{eqnarray*}}
\def\ese{\end{eqnarray*}}
\def\be{\begin{eqnarray}}
\def\ee{\end{eqnarray}}
\def\bq{\begin{equation}}
\def\eq{\end{equation}}
\def\bse{\begin{eqnarray*}}
\def\ese{\end{eqnarray*}}

\begin{document}
\pdfoutput=1
\thispagestyle{empty} \baselineskip=28pt \vskip 5mm
\begin{center} {\Huge{\bf Bayesian model averaging over tree-based dependence structures for multivariate extremes}}
\end{center}
\baselineskip=12pt \vskip 10mm

\begin{center}\large
Sabrina Vettori\footnote[1]{
\baselineskip=10pt King Abdullah University of Science and Technology (KAUST), Computer, Electrical and Mathematical Science and Engineering Division (CEMSE)
Thuwal 23955-6900, Saudi Arabia. \\ E-mails: sabrina.vettori@kaust.edu.sa, raphael.huser@kaust.edu.sa, marc.genton@kaust.edu.sa.},
Rapha\"el Huser$^1$,
Johan Segers\footnote[2]{
\baselineskip=10pt Universit\'{e} catholique de Louvain, Institut de Statistique, Biostatistique et Sciences Actuarielles (ISBA)
Louvain-la-Neuve B-1348, Belgium. \\ E-mail: johan.segers@uclouvain.be.}
and Marc G.~Genton$^1$
\end{center}
\baselineskip=17pt \vskip 10mm \centerline{\today} \vskip 10mm

\allowdisplaybreaks
\begin{center}
{\large{\bf Abstract}}
\end{center}

\baselineskip=17pt

\noindent Describing the complex dependence structure of extreme phenomena is particularly challenging. To tackle this issue we develop a novel statistical method that describes extremal dependence taking advantage of the inherent hierarchical dependence structure of the max-stable nested logistic distribution and that identifies possible clusters of extreme variables using reversible jump Markov chain Monte Carlo techniques. Parsimonious representations are achieved when clusters of extreme variables are found to be completely independent. Moreover, we significantly decrease the computational complexity of full likelihood inference by deriving a recursive formula for the nested logistic model likelihood. The algorithm performance is verified through extensive simulation experiments which also compare different likelihood procedures. The new methodology is used to investigate the dependence relationships between extreme concentrations of multiple pollutants in California and how these concentrations are related to extreme weather conditions. Overall, we show that our approach allows for the representation of complex extremal dependence structures and has valid applications in multivariate data analysis, such as air pollution monitoring, where it can guide policymaking.

\par\vfill\noindent
{\textbf{Keywords}: Air pollution; Extreme event; Fast likelihood inference; Nested logistic model; Reversible jump Markov chain Monte Carlo.} 
\par\medskip\noindent

\clearpage\pagebreak\newpage \pagenumbering{arabic}
\baselineskip=26pt

\section{Introduction}\label{sec: intro}
Estimating the probabilities associated with multivariate extreme phenomena beyond the original observation range requires flexible, yet interpretable, models with sound theoretical underpinning, that can be efficiently fitted to the data. However, these criteria are exponentially more difficult to meet as dimensionality increases; see, e.g., the review papers by \citet{Davison2012b} and \citet{Davison2015}. 
For these reasons the notion of ``high-dimensionality'' in the field of extremes typically invokes far smaller scales than those considered in standard statistics. 
Indeed, most applications of multivariate Extreme-Value Theory have focused chiefly on relatively low-dimensional cases, although much higher dimensions have been handled in structured, spatial, and spatio-temporal settings \citep{Wadsworth2012,Huser2014}. Moreover, the analysis of extreme events is also very challenging because of the intrinsic lack of data; extreme datasets comprise only the largest observations in a sample, and therefore are generally characterised by small sample sizes, i.e., few temporal replicates.

Recent literature has been focusing on developing innovative statistical methods to simplify the complex dependence structure of multivariate extremes in moderate or large dimensions.
\citet{Chautru2015} proposed a non-parametric technique able to identify possibly overlapping clusters of asymptotically dependent extreme variables and therefore reducing the dimension of the initial problem. 
Another approach, proposed by \citet{Bernard2013}, consists in a clustering technique for pointwise maxima data observed at different spatial locations; the new method integrates the $F$-madogram distance in the $K$-means algorithm. Moreover, \citet{Cooley2018} summarised tail dependence in high dimensions by the decomposition of a matrix of pairwise dependence metrics. 

In this work, we contribute to this developing research field by proposing an efficient methodology to describe parsimoniously the dependence between maxima of several variables. The technique can easily be adapted to threshold exceedances. 
Our approach relies on the nested logistic distribution \citep*{McFadden1978, Tawn1990, Coles1991,Stephenson2003}, which is max-stable and, therefore, asymptotically justified for the description of the whole multivariate extremal dependence structure \citep{DeHaan1977,DeHaan1984}. In particular, thanks to its hierarchical structure, the nested logistic model is able to describe the dependence within and between distinct clusters of extreme variables using logistic distributions \citep{Gumbel1960a,Gumbel1960b} within a tree-based representation. 

Likelihood-based inference for max-stable distributions is known to be challenging because of its computational burden. Indeed, the full likelihood in large dimensions becomes numerically intractable \citep{Castruccio.etal:2014, Bienvenue2014}. One classical solution that dramatically reduces the computational burden but leads to a loss of efficiency consists in using composite likelihood techniques \citep{Varin2005,Varin2008,Padoan2010, Genton2011, Davison2012,Huser2013,Castruccio.etal:2014}. The simplified likelihood procedure proposed by  \citet{Stephenson2005} leads to more efficient inference both computationally and statistically, but presents some bias at finite levels. Apart from its interpretability, the nested logistic model is appealing for its inference properties. Here, we derive a recursive formula for the likelihood of the nested logistic model. The formula greatly simplifies computations when dealing with multivariate extremes in moderate or high dimension without causing any efficiency loss or introducing any additional bias. 

Few authors have so far implemented a fully Bayesian inference method for the estimation of extremal dependence; see, e.g., \citet{guillotte:perron:segers:2011}, \citet{Ribatet2012},  \citet{Reich2012}, \citet{sabourin:naveau:fougeres:2013}, \citet{sabourin:naveau:2014}, \citet{Shaby2014} and \citet{Thibaud2016}. 
Our method is based on the reversible jump Markov chain Monte Carlo (RJ-MCMC) algorithm \citep{Green1995}, which allows us to estimate the nested logistic model parameters and simultaneously identify the most likely dependence structures governing the extremal dependence. Similarly to the Bayesian CART introduced by \citet{Chipman1998}, we implement the RJ-MCMC algorithm to identify the most homogeneous clusters of variables, represented within a tree framework. Moreover, instead of providing just one tree estimate, like in the classical CART method, our algorithm provides a natural measure of model uncertainty, which is given by the posterior probabilities calculated as the number of times a specific tree has been considered in the algorithm chain. Model uncertainty is taken into account for predictions using Bayesian model averaging, i.e., we construct the posterior predictive distributions as mixtures of the model-specific distributions weighted by the associated posterior probabilities. For more details on Bayesian model averaging see, e.g., \cite{Hoeting1999}.
Therefore, we are able to describe complex extremal dependence structures using a mixture of clustering configurations, transcending the partial exchangeability limitations of the nested logistic model while maintaining a moderate number of parameters. 
An extensive simulation study is conducted in order to evaluate the algorithm's performance, implementing both the \citet{Stephenson2005} likelihood and our recursive likelihood formula within the algorithm.
Finally, we apply our algorithm to multivariate air pollution data in order to investigate the dependence relations between extreme concentration of air pollutants and to understand how these relate to extreme weather conditions. 

The general framework of the Extreme-Value Theory and the nested logistic model are recalled in Section~\ref{sec: dependence} and the associated likelihood inference methods, including the proposed recursive likelihood formula, are summarised in Section~\ref{sec: known_inference}. In Section~\ref{sec: unknown_inference} we describe the proposed dimension reduction and clustering algorithm, and in Section~\ref{sec: simulation} we verify its performance through an extensive simulation study. The algorithm is applied to air quality data in Section~\ref{sec: application}. Finally, these results are discussed in Section~\ref{sec: discussion}. Theoretical details are reported in the Appendix and the Supplementary Material. 
\section{Modelling Extremal Dependence}\label{sec: dependence}
\subsection{General framework}
Suppose that $\mathbf{Y}_{i}=(Y_{i;1},\ldots,Y_{i;D})^\top$, $i=1,2,\ldots$, is a sequence of independent and identically distributed (i.i.d.) copies of the
 $D$-dimensional random vector $\mathbf{Y}$, representing $D$ variables of interest having a common distribution function (d.f.) $F$ with margins $F_{d}(y_d), d=1,\ldots,D$, and let $\mathbf{M}_{n} = (M_{n;1},\ldots,M_{n;D})^\top = \left(\underset{1\leq i\leq n}{\textrm{max}}Y_{i;1},\ldots, \underset{1\leq i\leq n}{\textrm{max}}Y_{i;D}\right)^\top$
denote the vector of multivariate componentwise maxima computed over $n$ observations.
We assume that, as $n\to\infty$, for some sequences of vectors $\mathbf{a}_{n}=(a_{n;1}, \ldots,a_{n;D})^\top\in \mathbb{R}^D_+$ and $\mathbf{b}_{n}=(b_{n;1}, \ldots,b_{n;D})^\top\in \mathbb{R}^D$, the renormalised vector of componentwise maxima $\mathbf{M}_{n}^{*}=\mathbf{a}_{n}^{-1}(\mathbf{M}_{n}-\mathbf{b}_{n})$ converges in distribution to the random vector $\mathbf{Z}$, with limiting $D$-dimensional extreme-value d.f. $G$ and non-degenerate margins; that is, the distribution $F$ belongs to the max-domain of attraction (MDA) of $G$. 
Upon marginal standardisation, we may assume unit-Fr\'{e}chet
margins, i.e., $G_{d}(z_d)=\exp(-1/z_d)$ for $z_d>0, d=1,\ldots,D$. 
The joint distribution of the random vector $\mathbf{Z}$ may be written as
\begin{equation}\label{eq: joint}
\textrm{P}(\mathbf{Z}\leq \mathbf{z})=G(\mathbf{z})=\exp\{-V(\mathbf{z})\},\quad \quad
V(\mathbf{z})=\int_{S_{D}}\underset{1\leq d\leq D}{\max}\left(\frac{\omega_{d}}{z_{d}}\right)\textrm{d}H(\boldsymbol{\omega}),\quad \quad  \mathbf{z} \in\mathbb{R}^D_+,
\end{equation}
where $V(\mathbf{z})$ is a homogeneous function of order $-1$, i.e., $V\left(s\mathbf{z}\right)=s^{-1}V\left(\mathbf{z}\right)$, called exponent function, and $H$ is a finite spectral measure on the unit simplex $S_{D}=\{\boldsymbol{\omega}\in [0,1]^{D} \colon \sum_{d=1}^D\omega_d=1\}$ for $\boldsymbol{\omega}=(\omega_{1},\ldots,\omega_{D})^\top$, satisfying the mean constraints $\int_{S_D} \omega_d \textrm{d}H(\boldsymbol{\omega})=1,d=1,\dots,D$ \citep{DeHaan1977,DeHaan1984}.

A useful summary of the dependence strength between multivariate extremes is the extremal coefficient, proposed by \citet{Smith1990} and defined as $\theta_D=V(\mathbf{1})\in [1,D],$ where $\mathbf{1}$ is a vector of ones. The coefficient $\theta_D$ decreases as the dependence strength between the margins increases, with $\theta_D=1$ and $\theta_D=D$ corresponding to complete dependence and independence respectively. 
\subsection{The logistic and nested logistic models}
Among max-stable distributions, the oldest and simplest one is the logistic model \citep{Gumbel1960a,Gumbel1960b}, characterised by the exponent function
\begin{equation}\label{eq: logistic}
V_l(\mathbf{z}\mid \alpha_0)=\left(\sum_{d=1}^D z_d^{-1/\alpha_0}\right)^{\alpha_0}, \quad \alpha_0\in(0,1].
\end{equation}
The parameter $\alpha_0$ summarises the dependence strength between the extreme observations $\mathbf{z}=(z_1,\ldots,z_D)^\top \in\mathbb{R}^D_+$. In particular, the cases of complete dependence and independence between the margins correspond to $\alpha_0\to0$ and $\alpha_0=1$, respectively. Therefore, the logistic model summarises the extremal dependence structure by only one parameter $\alpha_0$, and its components are exchangeable. 
The extremal coefficient for the logistic distribution is $\theta_D=D^{\alpha_0}$.

Generalising the idea of the logistic model, \citet{McFadden1978} and \citet{Tawn1990}, see also \citet{Coles1991}, proposed the nested logistic distribution, a more flexible multivariate max-stable model that maintains the logistic model's simplicity and interpretability. The nested logistic model implies that the vector $\mathbf{z}$, containing the extreme observations, is split into $K$ homogeneous clusters and it describes the dependence within and between these clusters using the logistic distributions defined in~\eqref{eq: logistic}. It is defined by the exponent function 
\begin{equation}\label{eq: nlogistic}
V_{nl}(\mathbf{z}\mid \boldsymbol{\alpha})=V_{l}\left\{V_{l}(\mathbf{z}_{1}\mid \alpha_0\alpha_1)^{-1},\ldots,V_{l}(\mathbf{z}_{K}\mid \alpha_0\alpha_K)^{-1}\mid \alpha_0\right\}=\left\{\sum_{k=1}^K \left(\sum_{i_k=1}^{D_k} z_{k;i_k}^{-\frac{1}{\alpha_0\alpha_k}}\right)^{\alpha_k}\right\}^{\alpha_0},
\end{equation}
where 
$V_{l}(\mathbf{z}_{k}\mid \alpha_0 \alpha_k)$ is the logistic exponent function in~\eqref{eq: logistic} of the sub-vector $\mathbf{z}_{k}=(z_{k;1},\ldots,z_{k;D_k})^\top$ comprising extreme observations belonging to the $k^\textrm{th}$ cluster of dimension $D_k \in  \{1,\ldots,D\}$, $k=1,\ldots,K,$ with $D=\sum_{k=1}^K D_k$, and where $\boldsymbol{\alpha}=(\alpha_0,\alpha_1,\ldots,\alpha_K)^\top\in(0,1]^{K+1}$ are the between-cluster and within-cluster dependence parameters. More precisely, the parameter $\alpha_0$ summarises the dependence strength between the clusters and the product of the parameters $\alpha_0\alpha_k$ summarises the dependence strength within the $k^\textrm{th}$ cluster. The extremal coefficient, which quantifies the effective number of independent variables among the $D$ variables, for the nested logistic distribution is $\theta_D=\left(\sum_{k=1}^K D_k^{\alpha_k}\right)^{\alpha_0}$. 
\begin{figure}[t!]
\hspace{6.2cm}{\centering\includegraphics[scale=0.48]{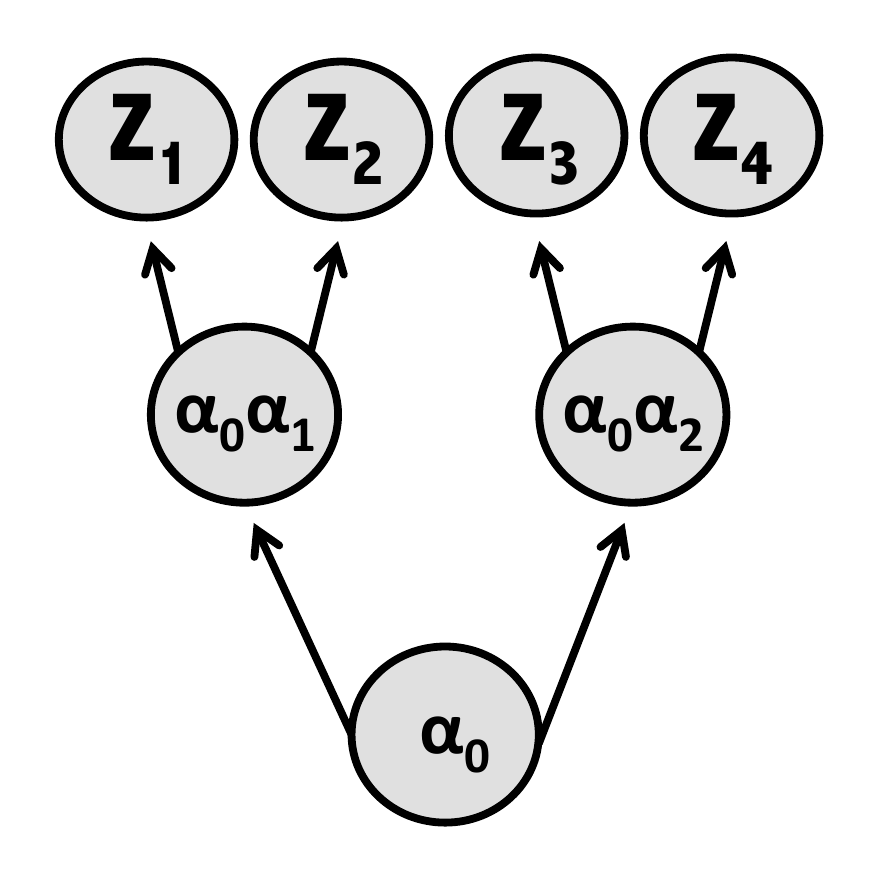}}
\caption{\footnotesize{Example of simple tree structure, summarising the extremal dependence of the vector $\mathbf{Z}=(Z_{1},Z_{2},Z_{3},Z_{4})^\top$, where the dependence between the variables $(Z_{i},Z_{j})^\top,$ $i\in\{1,2\},$ $j\in\{3,4\}$ is summarised by a logistic distribution with parameter $\alpha_0$, and the dependencies within the pairs of variables $(Z_{1},Z_{2})^\top$ and $(Z_{3},Z_{4})^\top$ are summarised by logistic distributions with parameters $\alpha_0 \alpha_1$ and $\alpha_0 \alpha_2$, respectively. }}\label{fig: tree.example}
\end{figure}
The hierarchical structure of the nested logistic model can be represented with a tree, as illustrated in Figure~\ref{fig: tree.example}. In practice, more complex situations may arise with more clusters and, possibly, an arbitrary number of layers. For simplicity, in this paper, we limit ourselves to dependence structures with only two layers.

The nested logistic model can also be defined in terms of nested Archimedean copulas or nested Gumbel copulas \citep{Okhrin2009,Hofert2013}. There exist already a number of papers on inference on such copulas, both parametric and non-parametric, see, e.g., \citet{okhrin:okhrin:schmid:2013}, \citet{segers:uyttendaele:2014} and \citet{gorecki:hofert:holena:2016}. 
\section{Full Likelihood Inference}\label{sec: known_inference}
\subsection{Classical approach}
The classical inference approach consists in splitting the observed data $\mathbf{Y}_{i}=(Y_{i;1},\ldots,Y_{i;D})^\top$, $i=1,\ldots, M$, into $N$ blocks with an equal number of observations $n$ (assuming that $M=Nn$), from which componentwise maxima data $\mathbf{m}_i=(m_{i;1},\ldots,m_{i;D})^\top,$ $i=1,\ldots,N,$ are extracted.
Assuming a multivariate extreme-value distribution with unit Fr\'{e}chet marginal distributions, the likelihood function for these vectors of maxima may be expressed as 
\begin{equation}\label{eq: l}
L(\boldsymbol{\alpha}\mid\mathbf{m}_1, \ldots, \mathbf{m}_N)=\prod_{i=1}^{N}\exp\{-V(\mathbf{m}_i\mid\boldsymbol{\alpha})\}\left\{\sum_{E\in \mathcal{E}}\prod_{S\in E}-\dot{V}_{S}(\mathbf{m}_i\mid\boldsymbol{\alpha})\right\},
\end{equation}
where $\boldsymbol{\alpha}$ is the vector of unknown dependence parameters, $\mathcal{E}$ denotes the collection of all partitions of $\mathcal{D}=\{1,\ldots,D\}$, $\dot{V}_{S}$ denotes the partial
derivative of the function $V$ in \eqref{eq: joint} with respect to the variables whose indices lie in $S \subset\mathcal{D}$.
The cardinality of $\mathcal{E}$ equals $B_D$, the Bell number of order $D$ \citep{Graham1988}, meaning that the number of terms to be computed and the storage space required for evaluating the likelihood grow super-exponentially with the dimension $D$ \citep{Castruccio.etal:2014, Huser2016}. 
\subsection{Recursive full likelihood formula}\label{sec: rec_inference}
Similarly to the recursive full likelihood formula for the logistic distribution that was derived by \citet{Shi1995}, we here develop a recursive formula to efficiently compute the full likelihood function~\eqref{eq: l} for the nested logistic distribution \eqref{eq: nlogistic}. 
Let $\boldsymbol{m}_{i;k;{1:D_k}}$ be the sub-vector of maxima belonging to the $k^\textrm{th}$ cluster of dimension $D_k$; the recursive likelihood formula may be written as
\begin{equation}\label{eq: density}
\begin{aligned}
&L(\boldsymbol{\alpha}\mid\mathbf{m}_1, \ldots, \mathbf{m}_N)= \prod_{i=1}^{N}\exp\{-V_{nl}(\mathbf{m}_{i}\mid\boldsymbol{\alpha})\} \prod_{i_1=1}^{D_1}m_{i;1;i_1}^{-\frac{1}{\alpha_0\alpha_1}-1}\cdots\prod_{i_K=1}^{D_K}m_{i;K;i_K}^{-\frac{1}{\alpha_0\alpha_k}-1}\sum_{i_{1}=1}^{D_{1}}\cdots\sum_{i_{K}=1}^{D_{K}}\\
&\times  \sum_{j=1}^{\sum_{k=1}^K i_k}\beta_{i_{1};\ldots;i_{K};j}^{(D_{1};\ldots;D_{K})}V_{l}(\mathbf{m}_{i;1;1:D_1}^{1/\alpha_0}\mid \alpha_1)^{i_{1}-\frac{D_{1}}{\alpha_1}}\cdots V_{l}(\mathbf{m}_{i;K;1:D_K}^{1/\alpha_0}\mid \alpha_k)^{i_{K}-\frac{D_{K}}{\alpha_k}}V_{nl}(\mathbf{m}_{i}\mid\boldsymbol{\alpha})^{j-\frac{\sum_{k=1}^K i_k}{\alpha_0}},
\end{aligned}
\end{equation}
where $V_{nl}(\mathbf{m}_i\mid \boldsymbol{\alpha})$ is defined in \eqref{eq: nlogistic} and $V_{l}(\mathbf{m}_{i;k;1:D_k}^{1/\alpha_0}\mid \alpha_k)=\left(\sum_{i_k=1}^{D_k} m_{i;k;i_k}^{-1/\alpha_0\alpha_k}\right)^{\alpha_k},$ $i=1, \ldots, N,$ $k=1,\ldots,K$. The coefficients $\beta_{j_{1};\ldots;j_{K};j}^{(D_1;\ldots;D_K)}$ can be computed recursively in explicit form reducing the computational complexity to 
$
\mathcal{O}\left(\sum^K_{k=1}\left(D_1+\cdots+D_k\right)D_1\cdots D_{k-1}D_k^2\right)\ll B_D;$
the proof of \eqref{eq: density} and the expression for $\beta_{j_{1};\ldots;j_{K};j}^{(D_1;\ldots;D_K)}$ are given in Appendix~\ref{sec:proof}. 
When all the clusters are of equal size, \textit{i.e.,} $D_k=D_1$ for all $k=2,\ldots,K$, the complexity is  $\mathcal{O}\left(K\sum_{k=1}^KD_1^{k+2}\right)$.  
Therefore, the computational time needed to compute the full likelihood using the recursive formula is polynomial in terms of the number of variables $D$, but exponential-linear in terms of the number of clusters $K$, suggesting that in practice it might be convenient to prevent $K$ from being large when $D$ is large. 
Recursive formulas for deeper trees with more layers may be derived similarly, although at a higher computational complexity.

\subsection{Stephenson and Tawn likelihood}
\citet{Stephenson2005} developed a simplified likelihood approach which uses the additional information of the partition representing the componentwise maxima that occurred simultaneously. More specifically, for each $i=1,\ldots,N$, let $E_i \in \mathcal{E}$ denote the partition grouping the maxima $m_{i;1},\ldots,m_{i;D}$ with identical occurrence times.  
For instance, for $D=3$, $E_1=\{1, \{2, 3\}\}$ indicates that, in the first block, the maxima $m_{1;2}$ and $m_{1;3}$ occurred simultaneously, but separately from $m_{1;1}$. 
The Stephenson and Tawn likelihood might be written as
\begin{equation}\label{eq: stl}
L\{\boldsymbol{\alpha}\mid(\mathbf{m}_1, E_1), \ldots, (\mathbf{m}_N, E_N)\}=\prod_{i=1}^{N}\exp\{-V(\mathbf{m}_{i}\mid\boldsymbol{\alpha})\}\left\{\prod_{S\in E_i}-\dot{V}_{S}(\mathbf{m}_{i}\mid\boldsymbol{\alpha})\right\},
\end{equation}
where $\boldsymbol{\alpha}$ is the vector of unknown parameters and $\dot{V}_{S}$ denotes the partial derivative of $V$ in \eqref{eq: joint} with respect to the variables in $S$. Therefore, using the partitions leads to a more efficient inference, both computationally and statistically, which comes at the price of introducing bias in the parameter estimation; see, e.g., \citet{Huser2016}, \citet{Thibaud2016} and our simulation experiment in Section \ref{sec: simulation}. Indeed, the Stephenson and Tawn likelihood conditions on the \emph{asymptotic} partitions being equal to the observed, \emph{sub-asymptotic} ones, which might be misspecified and cause bias for finite block sizes $n$. 
\citet{Wadsworth2014} proposed a bias reduction technique, which works well in low dependence scenarios, but remains intensive in case of strong dependence for large dimensions. 
To further speed up the computation of \eqref{eq: stl}, we derived a recursive formula to compute the partial derivatives of the nested logistic model exponent function. Let $\boldsymbol{m}_{i;k;{1:d_k}}$ be the sub-vector of the first $1\leq d_k\leq D_k$ components of the maxima vector belonging to the $k^\textrm{th}$ cluster. Partial derivatives of the nested logistic exponent function $V_{nl}$ with respect to the variables $\boldsymbol{m}_{i;k;{1:d_k}}$, $k=1,\ldots,\kappa$, within $1\leq\kappa\leq K$ clusters, may be expressed as
\begin{equation}\label{rec_stl}
\begin{aligned}
&\frac{\partial^{\sum_{k=1}^\kappa d_k}V_{nl}(\mathbf{m}_{i}\mid\boldsymbol{\alpha})}{\partial \prod_{k=1}^\kappa \mathbf{m}_{i;k;1:d_k}}=\prod_{i_1=1}^{d_1}m_{i;1;i_1}^{-\frac{1}{\alpha_0\alpha_1}-1}\cdots\prod_{i_\kappa=1}^{d_\kappa}m_{i;\kappa;i_\kappa}^{-\frac{1}{\alpha_0\alpha_\kappa}-1}\\
&\times\sum_{i_1=1}^{d_1}\cdots\sum_{i_\kappa=1}^{d_\kappa}\gamma_{i_{1};\ldots;i_{\kappa}}^{(d_1;\ldots;d_{\kappa})}V_{l}(\mathbf{m}_{i;1;1:d_1}^{1/\alpha_0}\mid \alpha_1)^{i_1-\frac{d_1}{\alpha_1}}\cdots  V_{l}(\mathbf{m}_{i;\kappa;1:d_\kappa}^{1/\alpha_0}\mid \alpha_\kappa)^{i_\kappa-\frac{d_\kappa}{\alpha_\kappa}}V_{nl}(\mathbf{m}_{i}\mid\boldsymbol{\alpha})^{1-\frac{\sum_{k=1}^\kappa i_k}{\alpha_0}};
\end{aligned}
\end{equation}
the total complexity for the computation of the recursive coefficients $\gamma_{i_{1};\ldots;i_{\kappa}}^{(d_{1};\ldots;d_{\kappa})}$ is reduced to $\mathcal{O}\left(\sum^\kappa_{k=1}d_1\cdots d_{k-1}d_k^2\right)$. If all clusters are of equal size, i.e., $d_k=d_1$, for all $k=2,\ldots,\kappa$, then one has $\mathcal{O}\left(\sum^\kappa_{k=1}d_1^{k+1}\right).$ The proof of \eqref{rec_stl} and the expression for $\gamma_{i_{1};\ldots;i_{\kappa}}^{(d_{1};\ldots;d_{\kappa})}$ are given in the Supplementary Material.

\section{Bayesian model averaging over tree-based dependence structures}\label{sec: unknown_inference}
\subsection{Posterior sampling of parameters using Metropolis-Hastings}
In this section, we introduce our novel inference approach to estimate the nested logistic model given a known tree structure, whereas in the following section we generalise it, in order to simultaneously identify the most likely tree structures representing the extremal dependence of the data. The Metropolis-Hastings Markov chain Monte Carlo (MH-MCMC) algorithm \citep{Hastings1970} is used to estimate the nested logistic model parameters $\boldsymbol{\alpha}=(\alpha_0, \alpha_1,\ldots,\alpha_K)^\top$. We sample from the posterior distribution $\pi(\boldsymbol{\alpha}\mid\mathbf{m}_1, \ldots, \mathbf{m}_N)$, obtained as a product of prior distributions $\pi(\boldsymbol{\alpha})=\pi(\alpha_0)\pi(\alpha_1)\cdots\pi(\alpha_K)$ and the likelihood $L(\boldsymbol{\alpha}\mid\mathbf{m}_1, \ldots, \mathbf{m}_N)$ in \eqref{eq: stl} using \eqref{rec_stl}, where the partition notation has been omitted for simplicity, or in \eqref{eq: density}. 
At each MH-MCMC iteration, a candidate parameter value $\alpha_p^*$ is proposed for each parameter $\alpha_p$ based on the proposal distribution $q( \alpha_p^*\mid\alpha_p^{c})$, where $\alpha_p^c$ denotes the current value of the parameter $\alpha_p$. To allow for independence within and between the clusters of variables, the dependence parameters in~\eqref{eq: nlogistic} has to take values at the boundary of their domain of definition, i.e., $\alpha_p=1$, $p=0, \ldots, K.$ Similarly to \citet{Coles2002}, we allow such flexibility by defining the proposal distribution as a mixture between a uniform distribution centered at the current value of the parameter $\alpha_p^{c}$ with an interval of length $2\varepsilon_p$, and a point mass at $\alpha_p=1$, i.e.,
\[
q( \alpha_p^*\mid\alpha_p^{c})=
\left\{ \begin{array}{ll}
 0.5\delta_1 + 0.5 p( \alpha_p^*\mid\alpha_p^{c}), & \textrm{if } \alpha_p^{c}\neq1 \; \textrm{and} \;1 \in [\alpha_p^{c}-\varepsilon_p, \alpha_p^{c}+\varepsilon_p];\\
p(\alpha_p^*\mid\alpha_p^{c}), & \textrm{if } \alpha_p^{c}=1\; \textrm{or} \; 1 \notin [\alpha_p^{c}-\varepsilon_p, \alpha_p^{c}+\varepsilon_p],
\end{array} \right.
\]
where $\delta_1$ denotes the Dirac delta function centered at 1 and 
$p( \alpha_p^*\mid\alpha_p^{c})$ is the density of a uniform random variable with boundaries $\left\{\max\left(0,\alpha_p^{c}-\varepsilon_p\right),\min\left(\alpha_p^{c}+\varepsilon_p,1\right)\right\}.$ The prior distribution of $\pi(\alpha_p)$ is defined as $0.5\delta_1 + 0.5\textrm{Unif}(0,1)$. 
At each iteration, the parameter $\alpha_p^{c}$ is updated based on the acceptance probability
$\min \left\{\frac{\pi(\boldsymbol{\alpha}^*\mid\mathbf{m}_1, \ldots, \mathbf{m}_N)q(\alpha_p^{c}\mid\alpha_p^*)}{\pi(\boldsymbol{\alpha}^{c}\mid\mathbf{m}_1, \ldots, \mathbf{m}_N)q(\alpha_p^*\mid\alpha_p^{c})},1\right\}$. Adaptive methodologies for choosing the
proposal tuning parameter $\varepsilon_p$ allow for efficient simulations even in high dimensions. We update $\varepsilon_p$ every 100 iterations during the burn-in, ensuring that the acceptance ratio takes values between 20\% and 50\%, in order to guarantee well-mixing properties of the chain. Then, we restart the algorithm with fixed $\varepsilon_p$ values. More details on the algorithm are available in the Supplementary Material. 
\subsection{Bayesian model averaging using reversible jump MCMC}
To estimate the unknown extremal dependence structure directly from the data, we explore an extension of the MH-MCMC algorithm, called reversible jump MCMC algorithm \citep{Green1995} (RJ-MCMC). Here the goal is to explore and identify the most likely clustering configurations for the components of the componentwise maxima vector. 
The nested logistic model representation implies that each partition of the componentwise maxima vector can be represented within a tree framework, see Figure \ref{fig: tree.example}, and therefore we refer to each clustering configuration as a tree structure. The reversible jump MCMC allows for the simulation from the posterior distribution on spaces of possibly varying dimensions and therefore it is capable of moving between the most likely tree structures representing the extremal dependence of the data. Bayesian model averaging provides a framework to take into account the uncertainty associated with the tree selection. In particular, the final posterior predictive distribution is obtained as an average of the posterior distributions associated with each of the trees considered by the algorithm weighted by their posterior probabilities, calculated as the proportion of time the chain has spent on each of the trees. 

Any proposed transition from the current tree $\mathcal T^c$, with parameters  $\boldsymbol{\alpha}^c$ of dimension dim$(\boldsymbol{\alpha}^c)$, to the proposed tree $\mathcal T^*$, with parameters $\boldsymbol{\alpha}^*$ of dimension dim$(\boldsymbol{\alpha}^*)$, must be reversible. Therefore, if the transition involves a dimension change, a random auxiliary variable $\mathbf{u}^c$ is generated such that, when moving from the current state $\mathbf{x}^c=(\boldsymbol{\alpha}^c, \mathbf{u}^c)$ to the proposed state $\mathbf{x}^*=(\boldsymbol{\alpha}^*, \mathbf{u}^*)$, the dimension matching condition $\textrm{dim}(\mathbf{x}^c)=\textrm{dim}(\mathbf{x}^*)$ is satisfied and the mapping $g_{\mathbf{x}^c\rightarrow\mathbf{x}^*}$ is a bijection \citep{Green1995}. Throughout, we assume a uniform prior distribution on the space of valid two-layer tree structures, although a more efficient algorithm with more informative priors might be designed. 
The proposed transition is accepted based on the acceptance probability
  \[
\min \left\{\frac{\pi(\boldsymbol{\alpha}^*\mid\mathbf{m}_1, \ldots, \mathbf{m}_N)}{\pi(\boldsymbol{\alpha}^c\mid\mathbf{m}_1, \ldots, \mathbf{m}_N)}\frac{q(\mathbf{u}^{c},\boldsymbol{\alpha}^{c}\mid\mathbf{u}^{*},\boldsymbol{\alpha}^{*})}{q(\mathbf{u}^{*},\boldsymbol{\alpha}^{*}\mid\mathbf{u}^{c},\boldsymbol{\alpha}^{c})}\frac{\pi_{\mathbf{x}^c\rightarrow\mathbf{x}^*}}{\pi_{\mathbf{x}^*\rightarrow\mathbf{x}^c}}\left|\frac{\partial (\boldsymbol{\alpha}^{*},\mathbf{u}^{*})}{\partial(\boldsymbol{\alpha}^{c},\mathbf{u}^{c})}\right|,1\right\},
\]
where $\mathbf{u}^*$ is the auxiliary variable generated to meet the reversibility condition, $\pi(\boldsymbol{\alpha}^*\mid\mathbf{m}_1, \ldots, \mathbf{m}_N)$ indicates the posterior distribution evaluated at the state $\mathbf{x}^*$ (conditional on the tree $\mathcal T^*$), $q(\mathbf{u}^{*},\boldsymbol{\alpha}^{*}\mid\mathbf{u}^{c},\boldsymbol{\alpha}^{c})$ is the proposal distribution for moving from the state $\mathbf{x}^c$ to the state $\mathbf{x}^*$ and $\pi_{\mathbf{x}^c\rightarrow\mathbf{x}^*}$  is the probability of choosing such a move type (i.e., from the current tree $\mathcal T^c$ to the proposed tree $\mathcal T^*$); $\mathbf{u}^c$, $\pi(\boldsymbol{\alpha}^c\mid\mathbf{m}_1, \ldots, \mathbf{m}_N)$, $q(\mathbf{u}^{c},\boldsymbol{\alpha}^{c}\mid\mathbf{u}^{*},\boldsymbol{\alpha}^{*})$ and $\pi_{\mathbf{x}^*\rightarrow\mathbf{x}^c}$  correspond to the reverse move counterparts and $\left|\frac{\partial (\boldsymbol{\alpha}^{*},\mathbf{u}^{*})}{\partial(\boldsymbol{\alpha}^{c},\mathbf{u}^{c})}\right|$ is the Jacobian of the transformation $g_{\mathbf{x}^c\rightarrow\mathbf{x}^*}$. In practice, there might be no need to generate any auxiliary variable in one direction or the other.   

\begin{figure}[t!]
{\centering\includegraphics[scale=0.34]{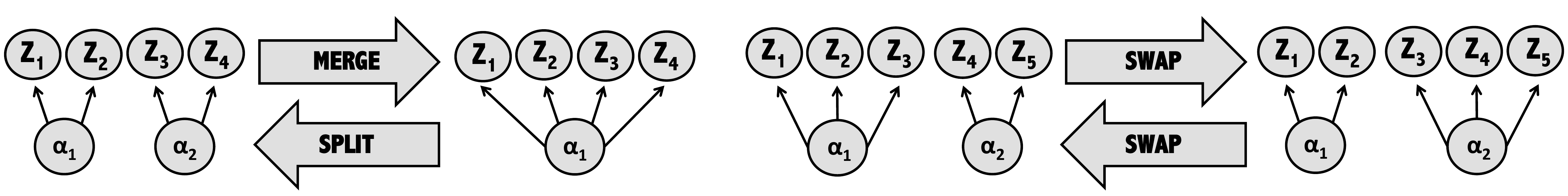}}
\caption{\footnotesize{Illustration of the reversible split, merge and swap moves implemented in the TM-MCMC algorithm.}}\label{fig: swap}
\end{figure}

We implement three different move types within the reversible jump MCMC algorithm in order to explore the space of trees. The moves are defined below and represented in Figure~\ref{fig: swap}. 
At each iteration of the reversible jump MCMC algorithm, each move type is chosen with probability $\pi_{\mathbf{x}^c\rightarrow\mathbf{x}^*}=1/3$. Given 
that our approach aims at describing extremal dependence using a mixture of tree structures, we henceforth refer to this as the Tree Mixture (TM)-MCMC algorithm.

\paragraph{Split Move}
The split move consists in splitting an existing cluster at random. It is the inverse of the merge move defined below, and it implies a dimension change transition. More precisely, the current state  $\mathbf{x}^{c}=(\alpha_1^c, u^c)^\top$ is mapped to the proposed state as $\mathbf{x}^{*}=(\alpha^*_1=\alpha_1^c+u^c, \alpha^*_{2}=\alpha_1^c-u^c)^\top$, where $u^c$ is an auxiliary variable. The proposal distribution for the split move $q(  u^{c},\alpha_1^c\mid\alpha^*_1, \alpha^*_2)$ is defined by assuming that $u^c$ is a uniform random variable with boundaries chosen such that $\left\{\max\left(0,\alpha_1^{c}-\eta\right)\leq\alpha_1^*,\alpha_2^* \leq\min\left(\alpha_1^{c}+\eta,1\right)\right\}$, with $\eta \in [0,1]$. The constant $\eta$ should, in practice, be chosen in order to ensure that the chain is well-mixing; in our case we fix $\eta=0.4$. Here, the Jacobian reduces to $\left|\frac{\partial (\alpha_1^*,\alpha_2^*)}{\partial(\alpha_1^c, u^c)}\right|=2$.
\paragraph{Merge Move}
The merge move consists in merging two clusters at random. It is the inverse of the split move, implying a dimension change transition from the state $\mathbf{x}^{c}=(\alpha_1^c, \alpha_{2}^c)^\top$ to $\mathbf{x}^{*}=\left(\alpha_1^*=\frac{\alpha_1^c+ \alpha^c_{2}}{2}, u^*=\frac{\alpha_1^c- \alpha^c_{2}}{2}\right)^\top$. In this case there is no need to generate any auxiliary random variable and the Jacobian reduces to $\left|\frac{\partial (\alpha_1^{*},u^{*})}{\partial(\alpha_1^{c},\alpha_2^{c})}\right|=\frac{1}{2}$.
\paragraph{Swap Move}
The swap move consists in exchanging some variables from one cluster to another cluster at random. This move type is self-reversible because transitioning from the current state $\mathbf{x}^{c}=(\alpha_1^c, \alpha_{2}^c)^\top$ to the proposed state $\mathbf{x}^{*}=(\alpha_1^*, \alpha_{2}^*)^\top$ does not involve any parameter dimension change and therefore the Jacobian is simply equal to one.

In our implementation of this algorithm, the initial tree configuration groups all variables in the same cluster, so that the order in which the variables appear in the dataset does not matter. More details on the algorithm are available in the Supplementary Material.


\section{Numerical Experiments}  \label{sec: simulation}
\subsection{Simulation settings} 
\begin{figure}[t!]
\hspace{2.5cm}{\centering\includegraphics[scale=0.4]{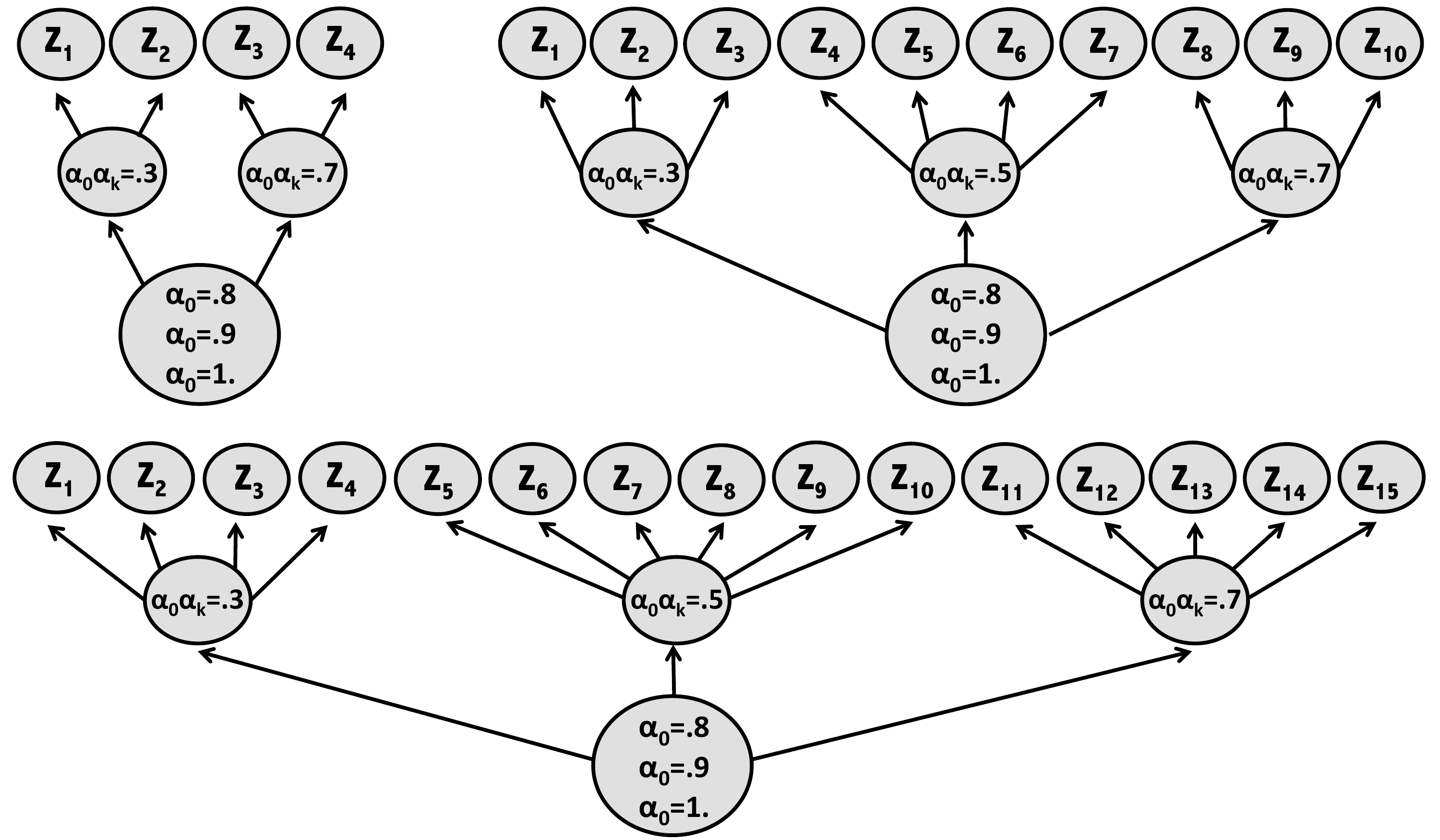}}
\caption{\footnotesize{Dependence structure configurations used for the data simulation when $D=4$ (top left), $D=10$ (top right) and $D=15$ (bottom).}}\label{fig: configuration}
\end{figure}

The TM-MCMC algorithm performance is tested on $1000$ independent simulated datasets of various dimensions $D = 4, 10, 15,$ imposing the dependence structures represented in Figure~\ref{fig: configuration}. It is important to bear in mind that multivariate extremes applications are normally carried out for relatively low dimensions. Here we simulate data up to dimensions $D=15$, which is considered fairly large in this type of applications. The data are generated from: the nested logistic distribution~\eqref{eq: nlogistic}, as explained by \cite{Stephenson2003}; a nested Archimedean copula with unit Fr\'{e}chet margins whose distribution is in the max-domain of attraction (MDA) of the nested logistic distribution; and the Student-$t$ copula with unit Fr\'{e}chet margins, whose limit max-stable distribution is not the nested logistic model, but rather the extremal $t$ model \citep{Opitz2013}, thus implying a more severe model misspecification. 
More specifically, data in the MDA of the nested logistic distribution are sampled using the outer power nested Clayton copula with two nesting levels and $K$ child copulas (clusters), i.e., 
\begin{equation}\label{eq: copula}
C(\mathbf{u}) = C\{C(\mathbf{u}_{1};\varphi_1),\ldots, C(\mathbf{u}_{K};\varphi_K);\varphi_0\},
\end{equation}
where $\mathbf{u}=(\mathbf{u}_1^\top,\ldots, \mathbf{u}_K^\top)^\top \in [0,1]^D$ and $\varphi_{p}=(1-t)^{-1/\theta_p},  p= {0,1, \ldots, K,}$ are Archimedean generators of the Clayton's family \citep{Hofert2013}. The parameters $\theta_p \in [1,\infty)$ fulfil the sufficient nesting condition if $\theta_0\leq \theta_k,$ $k=1, \ldots, K$ and are fixed according to $\theta_0=1/\alpha_0$ and $\theta_k=1/(\alpha_0 \alpha_k)$ with $\alpha_0, \alpha_k, k=1,2,3$, specified in Figure~\ref{fig: configuration}. Sampling from such copulas  is explained by \cite{Hofert2011}. More details about sampling nested Archimedean copulas using \verb R  can be found in \cite{Hofert2011a}. 
Student-$t$ data are generated from the multivariate Student-$t$ distribution with $10$ degrees of freedom, zero mean vector and covariance matrix $\boldsymbol{\Sigma}_{D\times D}(\rho_0)$ as specified below:
  \[\arraycolsep=2.5pt\def\arraystretch{1}
\boldsymbol{\Sigma}_{4\times4}(\rho_0)=\left[\begin{array}{cc}
\textbf{A}_{2}(\rho_1) &  \rho_0\textbf{1}_{2\times2}\\
 \rho_0\textbf{1}_{2\times2} & \textbf{A}_{2}(\rho_3)\\
\end{array}\right],
\]
\[\hspace{0.3cm}
\boldsymbol{\Sigma}_{10\times10}(\rho_0)=\left[\begin{array}{ccc}
\textbf{A}_{3}(\rho_1) & \rho_0\textbf{1}_{3\times4} & \rho_0\textbf{1}_{3\times3}\\
 \rho_0\textbf{1}_{4\times3} & \textbf{A}_{4}(\rho_2) &  \rho_0\textbf{1}_{4\times3}\\
 \rho_0\textbf{1}_{3\times3} &  \rho_0\textbf{1}_{3\times4} & \textbf{A}_{3}(\rho_3)
\end{array}\right],
\boldsymbol{\Sigma}_{15\times15}(\rho_0)=\left[\begin{array}{ccc}
\textbf{A}_{4}(\rho_1) & \rho_0\textbf{1}_{4\times6} & \rho_0\textbf{1}_{4\times5}\\
 \rho_0\textbf{1}_{6\times4} & \textbf{A}_{6}(\rho_2) &  \rho_0\textbf{1}_{6\times5}\\
 \rho_0\textbf{1}_{5\times4} &  \rho_0\textbf{1}_{5\times6} & \textbf{A}_{5}(\rho_3)
\end{array}\right],
\]
where $\textbf{1}_{n\times m}$ is the $n \times m$ matrix of ones and $\textbf{A}_n(\rho_k)=(1-\rho_k) \textbf{I}_n + \rho_k \textbf{1}_{n\times m}, k=1,2,3$, with $\textbf{I}_n$ being the $n \times n$ identify matrix and $\rho_k=0.98, 0.94, 0.86$. The simulation is repeated for different values of $\rho_0=0.77, 0.62, 0$. The coefficients $\rho_0$ and $\rho_k,$ $k=1,2,3$, are chosen such that the pairwise extremal coefficients $\theta_2$ for the limiting extremal $t$ distribution match the extremal coefficients calculated for the nested logistic model, with respective parameters $\alpha_0$ and $\alpha_0\alpha_k$, $k=1,2,3$, specified in Figure~\ref{fig: configuration}, except for $\rho_0=0$. Indeed, for $\rho_0=0$, $\theta_2=1.99$, only approximately $2$, which corresponds to the complete independence case of $\alpha_0=1$. The simulation parameters used in the numerical experiments are summarised in Table \ref{tab:est1}. For the results presented below in Section~\ref{sec:resultssimulation}, we identify for simplicity the Student-$t$ correlation parameters, $\rho_0,\rho_k$, to their analogue in terms of the nested logistic distribution, $\alpha_0,\alpha_0\alpha_k$.

\begin{table}[t!]
\caption{\footnotesize{Dependence parameters used in the simulation experiments. From left to right, the coefficients of $\alpha_0$ and $\rho_0$ represent the cases from weak dependence to complete or near independence between the clusters, and the coefficients $\alpha_0\alpha_k$ and $\rho_k$  represent the cases of strong, mild and weak dependence within the clusters, respectively. }}\label{tab:est1}
\begin{center}
\scalebox{0.78}{
\begin{tabular}{l|cccc|cccc}
 & \multicolumn{4}{c|}{MDA data} &  \multicolumn{4}{c}{Student-$t$ data} \tabularnewline
\midrule
$D=4$& $\alpha_0=0.8, 0.9, 1$ & $\alpha_0\alpha_k=0.3$ &$\alpha_0\alpha_k=0.7$ & & $\rho_0=0.77, 0.62, 0$ & $\rho_k=0.98$ &$\rho_k=0.86$ & \tabularnewline
$D=10$& $\alpha_0=0.8, 0.9, 1$ & $\alpha_0\alpha_k=0.3$ & $\alpha_0\alpha_k=0.5$ &$\alpha_0\alpha_k=0.7$ & $\rho_0=0.77, 0.62, 0$ & $\rho_k=0.98$ & $\rho_k=0.94$ &$\rho_k=0.86$ \tabularnewline
$D=15$& $\alpha_0=0.8, 0.9, 1$ & $\alpha_0\alpha_k=0.3$ & $\alpha_0\alpha_k=0.5$ &$\alpha_0\alpha_k=0.7$ & $\rho_0=0.77, 0.62, 0$ & $\rho_k=0.98$ & $\rho_k=0.94$ &$\rho_k=0.86$ \tabularnewline
\end{tabular}}
\end{center}
\end{table}

For each simulation setting under model misspecification we generate datasets of sample sizes $M=10000, 20000,40000$ and extract $N=100, 200,400$ componentwise maxima with blocks of size $n=100$, respectively.
The simulation results are obtained considering $R=15000$ algorithm iterations, including $R/5=3000$ burn-in iterations. We do not show simulation results for data sampled from the nested logistic distribution~\eqref{eq: nlogistic} as they lead to the same conclusion as simulation experiments conducted for data simulated from the nested outer power Clayton copula \eqref{eq: copula}.
\subsection{Summary of simulation results}\label{sec:resultssimulation}
\begin{figure}[t!]
\includegraphics[scale=0.256]{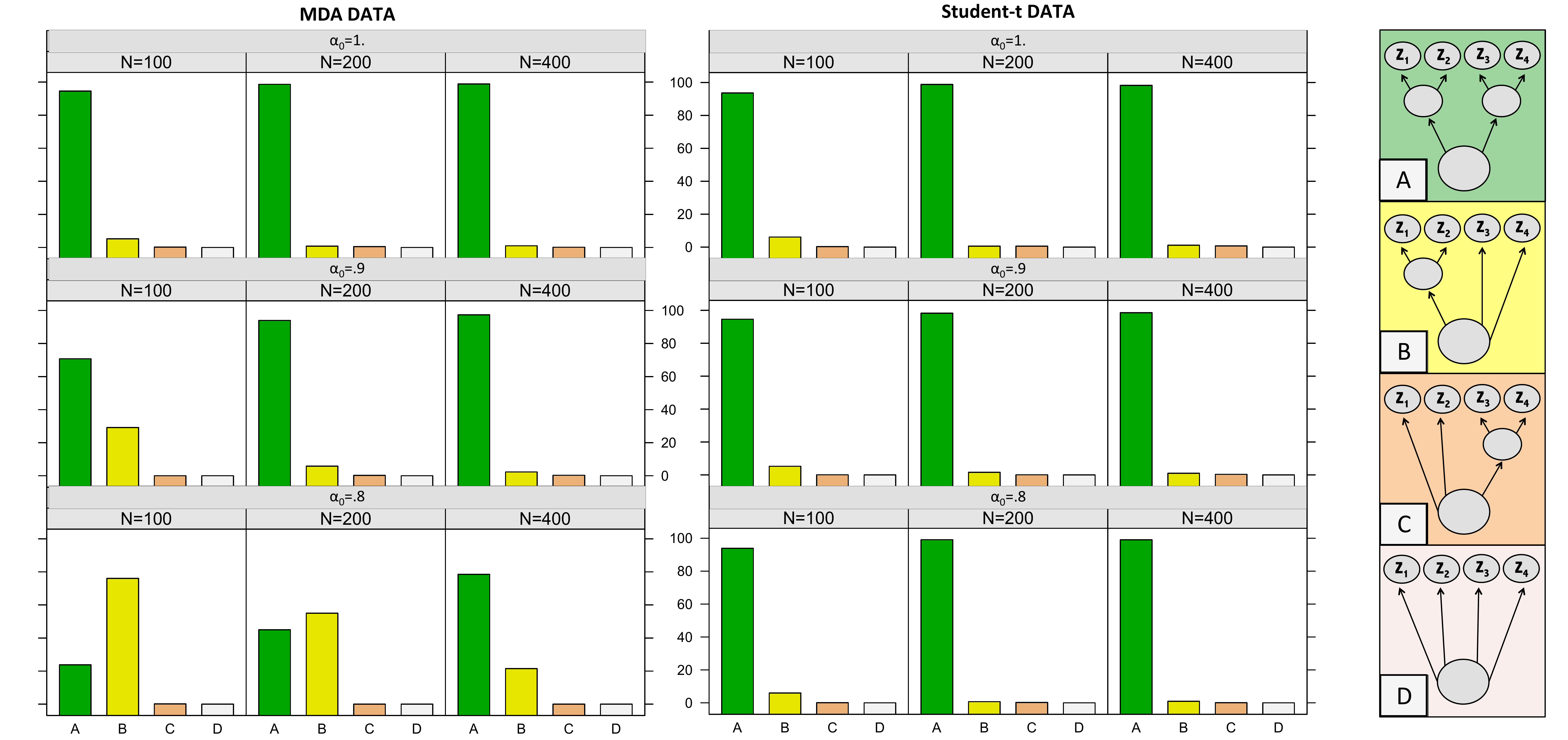}
\caption{\footnotesize{The histograms (left) report the proportion of times that a specific tree structure (right) is identified as the most likely one across $B=1000$ TM-MCMC algorithm chains after $R=15000$ iterations for data generated from the copula in~\eqref{eq: copula} and from the Student-$t$ copula with $D=4$ and tree structure comprising two clusters of variables (see the top-left tree in Figure~\ref{fig: configuration}). The recursive formula~\eqref{eq: density} was used for the likelihood evaluation. The correct tree structure is tree A (green).}}\label{fig: simulation1}
\end{figure}

\begin{figure}[t!]
{\centering\includegraphics[scale=0.6]{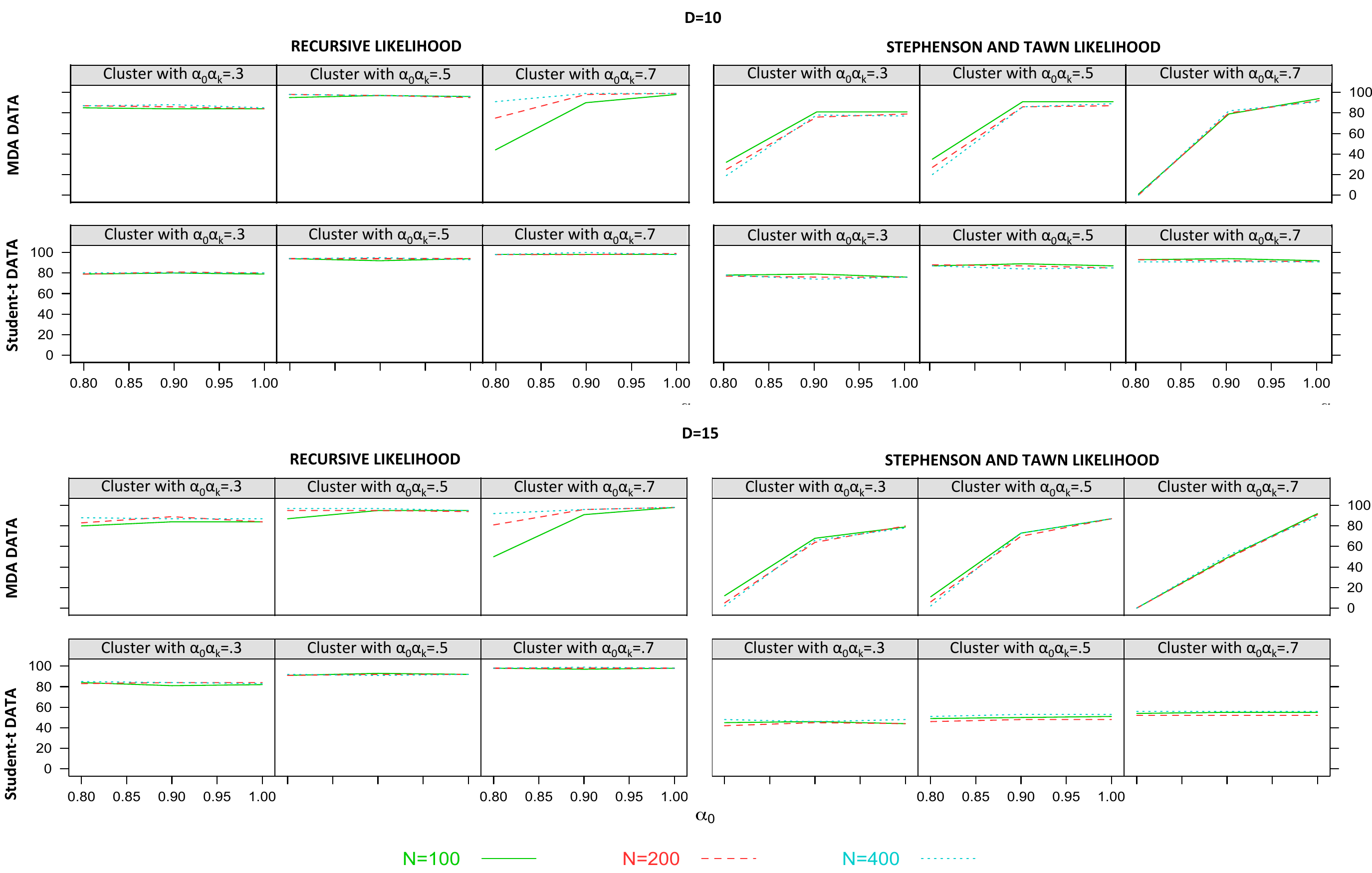}}
\caption{\footnotesize{The true positive rate calculated for each cluster is plotted against different values of the parameter $\alpha_0$ (abscissa) for different sample sizes $N$ (colours) obtained by applying the TM-MCMC algorithm with $R=15000$ iterations over $B=1000$ replicates for data generated from the copula in~\eqref{eq: copula} and from the Student-$t$ copula with $D=10,15$ and tree structure comprising three clusters of variables (see the bottom tree in Figure~\ref{fig: configuration}) implementing either the recursive formula~\eqref{eq: density} (left) or the Stephenson and Tawn~\eqref{eq: stl} likelihood (right) using~\eqref{rec_stl}, respectively.
}}\label{fig: simulation3}
\end{figure}
The histograms in Figure~\ref{fig: simulation1} indicate the proportion of times that each tree structure (A, B, C and D) is identified as the most likely clustering configuration representing the data across $B=1000$ TM-MCMC algorithm chains, when the recursive likelihood formula~\eqref{eq: density} is used in each likelihood evaluation, considering data generated from both the copula in~\eqref{eq: copula} and the Student-$t$ distribution. 
Generally, the algorithm correctly identifies the true tree structure, represented by tree A (green), as the most likely partition of componentwise maxima describing the data dependence structure. Moreover, the algorithm identifies at least the cluster characterised by strong dependence strength in tree B (yellow) across all chains. When the data are generated from the copula \eqref{eq: copula}, the algorithm performance improves as the value of the $\alpha_0$ parameter and the sample size increase; meaning that the between-cluster dependence strength plays an important role in identifying the true dependence structure. When data are simulated from the Student-$t$ distribution, the ``true'' tree structure is mostly identified by the algorithm for any sample sizes, suggesting that our algorithm might work well in misspecified settings.

Figure~\ref{fig: simulation3} displays the proportion of times the true componentwise maxima partition was identified by applying the TM-MCMC algorithm, also called true positive rates, when using either the recursive formula~\eqref{eq: density} or the Stephenson and Tawn likelihood~\eqref{eq: stl}. The true positive rates are plotted against different values of the parameter $\alpha_0$ for sample sizes $N=100, 200, 400$. The data are simulated from the copula~\eqref{eq: copula} and from the Student-$t$ distribution with $D=10,15$. In accordance with the findings for $D=4$, the algorithm performs reasonably well, identifying all clusters more than 80\% of the time for all sample sizes when using the recursive likelihood. 
The effect of misspecification is more apparent in larger dimensions, but remains quite weak overall. 
In contrast, the true positive rate is generally lower for all clusters when the Stephenson and Tawn likelihood is implemented, and it decreases as the between-cluster dependence increases. 

The computational time generally ranges from a few seconds to a few minutes depending on the running time for individual likelihood calculations.  In the simplest case with $D=4$ and $N=100$, a single likelihood evaluation takes around 0.05 seconds using either the recursive likelihood formula or the Stephenson and Tawn likelihood formula, whereas in the most complicated case with $D=15$ and $N=400$ a single likelihood evaluation takes around 12 seconds using the recursive likelihood and less than 2 seconds using the approximate Stephenson and Tawn likelihood. 
\section{Application to Air Pollution Data} \label{sec: application}
\subsection{Air pollutants} 
Air pollution has various negative effects on human health, ranging from respiratory illnesses to premature death, see, e.g., \citet{Peden2001}, \citet{Brunekreef2002} and \citet{Kampa2008}, and causes serious global environmental issues such as global warming and ozone depletion, see, e.g., \citet{Murphy1999} and \citet{Seinfeld2016}.  
Generally, regional air quality is affected by topography and weather, but also by emission sources.  
\begin{figure}[t!]
\begin{center}
\includegraphics[scale=0.56]{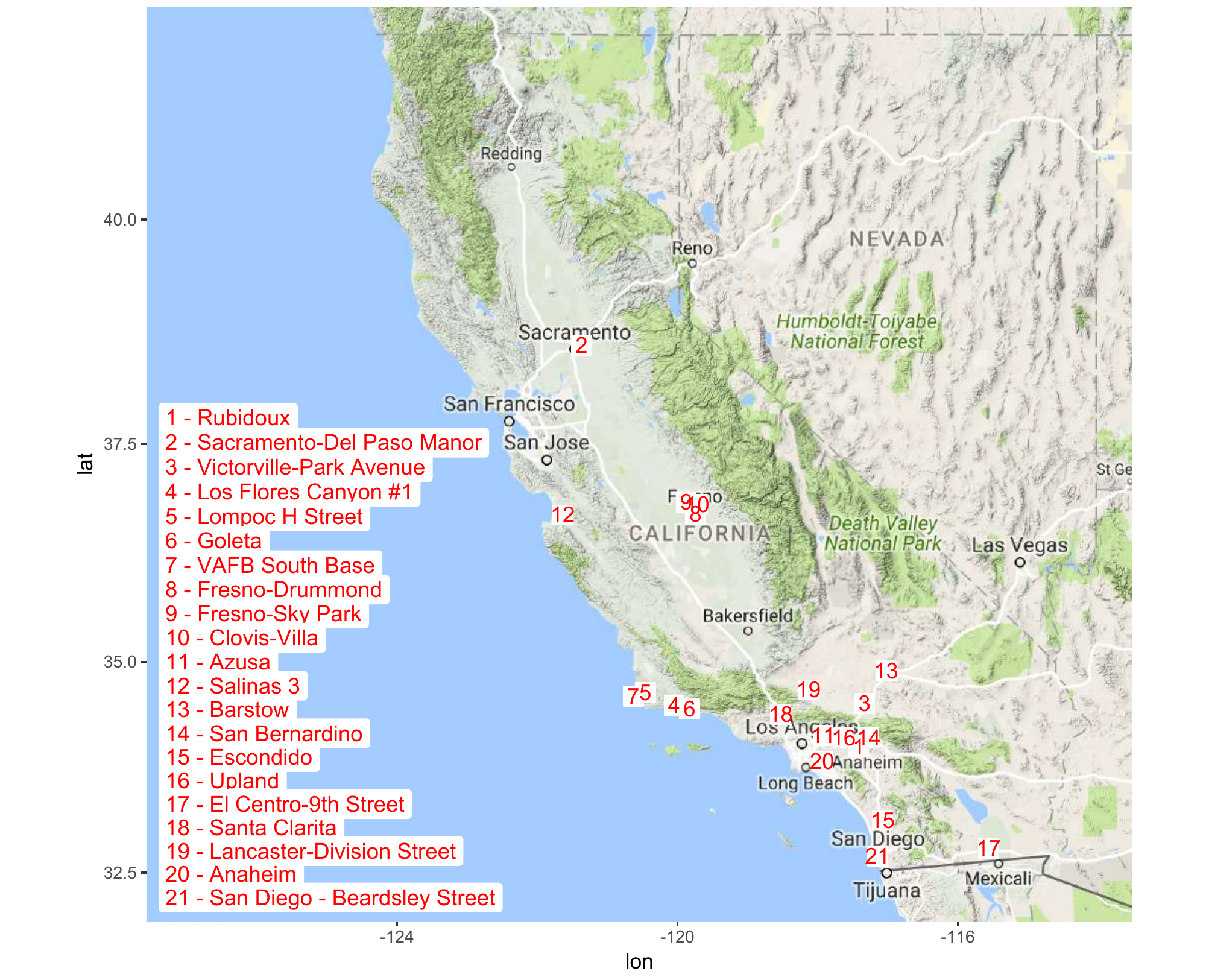}
\caption{\footnotesize{Map of California with the 21 sites under study indicated by numbers.
}}\label{fig: map}
\end{center}
\end{figure}
Ground level ozone (O$_3$) is formed by chemical reactions of volatile organic compounds (VOCs) and nitrogen dioxide (NO$_2$), in the presence of heat and sunlight \citep{Jacob2009}. NO$_2$, like its precursor nitric oxide (NO), and carbon monoxide (CO) are mostly created by the combustion of fossil fuels in power plants and automobile engines \citep{Cofala2007}. VOCs include a large variety of both natural and artificial chemical species, such as methane and isoprene. 

Several international institutions and agencies, including the United States Environmental Protection Agency (U.S. EPA), have traditionally focused on controlling the emissions of each of the most dangerous air pollutants, also called criteria pollutants, e.g., O$_3$, NO$_2$ and CO, separately. 
However, long-term and peak exposures to multiple air pollutants simultaneously have been demonstrated to cause serious health problems \citep{Dominici2010,Johns2012} and so policy-makers worldwide, including the U.S. EPA, are now keen on moving towards a multi-pollutant approach to quantify air pollution risks \citep{Johns2012}. There is also growing interest in studying the health effects of the interaction between ozone and temperature, see, e.g., \cite{Kahle2015}, particularly given that temperatures are expected to rise in the coming decades. In this work, we investigate the dependence relationships between extreme concentrations of air pollutants and extreme weather conditions through max-stable distributions using the TM-MCMC algorithm described in the previous sections. In particular, we focus on daily observations available from the Air Quality System (AQS) database on the EPA website \url{http://aqsdr1.epa.gov/aqsweb/aqstmp/airdata/download_files.html} collected at 21 sites across the state of California, which is one of the most populated and polluted areas of the US \citep{Air2017}. The site locations can be visualised in Figure~\ref{fig: map}. 
Details on the data preprocessing and marginal estimations are available in the Supplementary Material. 

\subsection{Multivariate analysis of California air pollution extremes}
\begin{figure}[t!]
{\centering\includegraphics[scale=0.32]{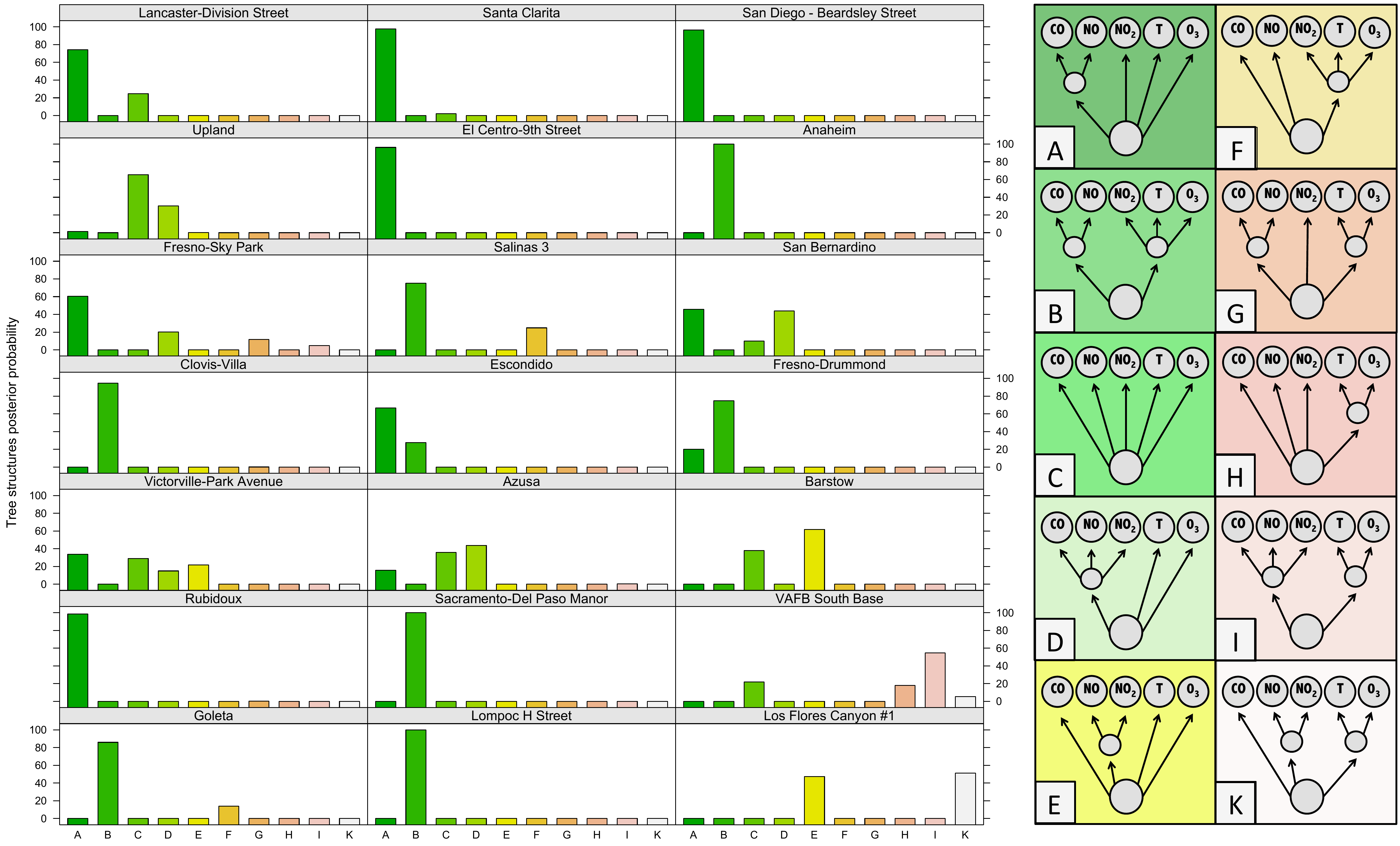}}
\caption{\footnotesize{A different letter is associated to each of the most frequent tree structures (right) identified by the TM-MCMC algorithm after $R=15000$ iterations and burn-in $R/5=3000$. The histograms (left) report the posterior probability associated to each tree for each site under study calculated according to the number of times each tree appears in the algorithm chain.
}}\label{fig: TREE}
\end{figure}
Considering daily observations of the variables CO, NO, NO$_2$, O$_3$ and temperature (T) collected from January 2004 to December 2015, we derive at least 100 multivariate monthly maxima for each site in Figure~\ref{fig: map} to which we apply the TM-MCMC algorithm with $R=15000$ iterations, and burn-in $R/5=3000$. The most frequent dependence structures identified by the algorithm and the corresponding posterior probabilities obtained for each of the sites under study are represented in Figure~\ref{fig: TREE}. 
In particular, the tree structures that appear most often across the chains are trees A and B, considered at 16 sites. Tree A includes only the CO-NO cluster, while tree B has two clusters, respectively grouping the maxima of NO and CO and the maxima of NO$_2$ and O$_3$ with T.  
While the data dependence structure can be represented by only tree B at the Clovis-Villa site, for instance the TM-MCMC algorithm suggests to represent the extremal dependence structure at Victorville-Park Avenue using a mixture of the trees A, C, D and E.

\begin{figure}[t!]
{\centering\includegraphics[scale=0.45]{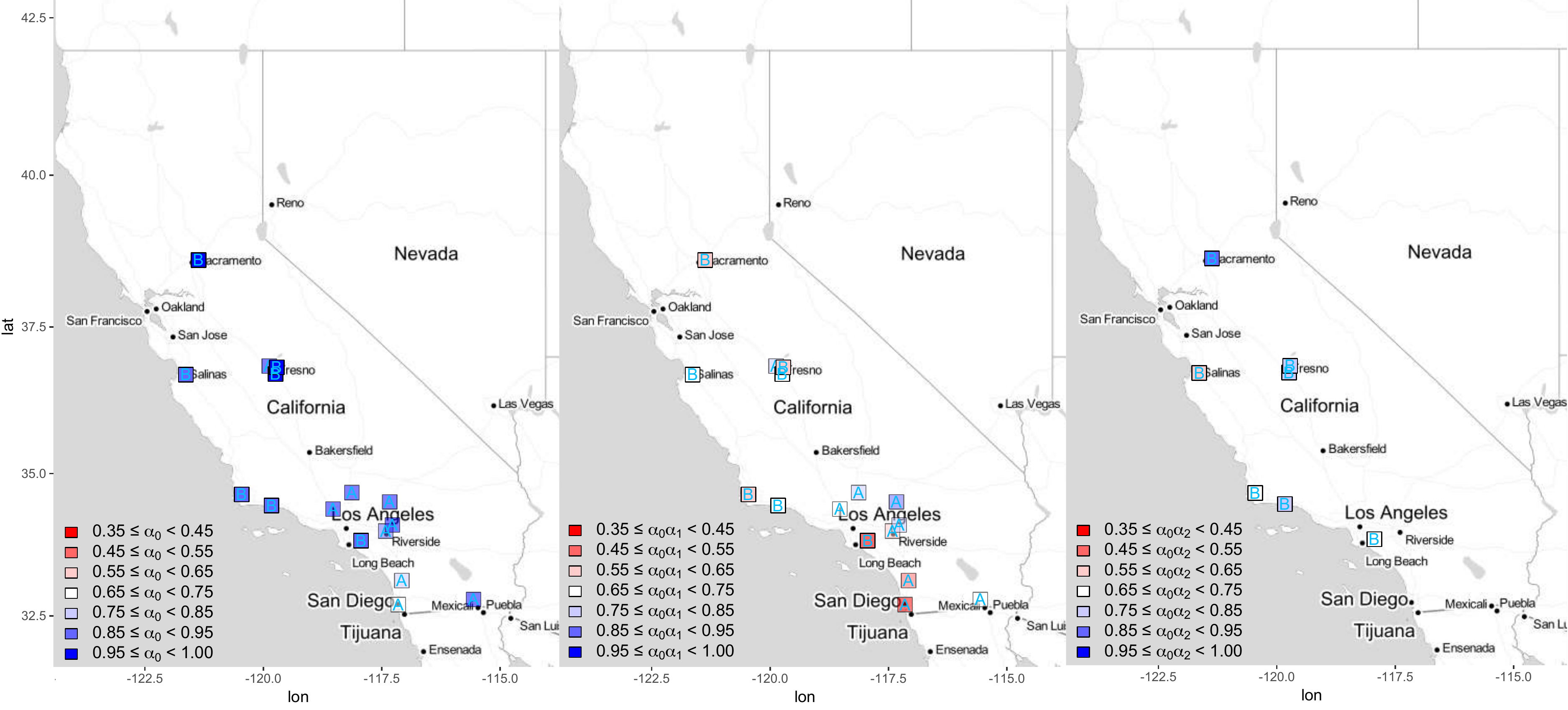}}
\caption{\footnotesize{A colour scale is used to represent the point estimates of the nested logistic model~\eqref{eq: nlogistic} parameters computed as the median of the sub-chain corresponding to the most likely tree after burn-in. The parameter $\alpha_0$ represents the dependence between the clusters (left map) and the parameters $\alpha_0\alpha_1$, $\alpha_0\alpha_2$ represent the dependence within the clusters CO-NO (central map) and O$_3$-NO$_2$-T (right map), respectively. }}\label{fig: dependence}
\end{figure}
Figure~\ref{fig: dependence} represents the point estimates for the nested logistic model parameters $\boldsymbol{\alpha}$, considering the tree structures A and B in Figure~\ref{fig: TREE}. 
The values of the parameter $\alpha_0$ represented in the left map are generally close to $\alpha_0=0.7$ for the southern sites, whereas $\alpha_0$ tends to be close to $1$ for the northern sites, indicating that most of the identified clusters can be treated independently. This suggests that the original observations from these distinct clusters are asymptotically independent, which can arise for example when they are weakly dependent in the bulk, but not in the upper tail, or when they are negatively associated. As expected, the dependence strength within the clusters is stronger. The maxima of NO and CO are found to be particularly dependent in areas characterised by heavy traffic, especially between the cities of Anaheim and Long Beach and close to Los Angeles and San Diego. The dependence strength between the maxima of NO$_2$, O$_3$ and T is generally mild or weak for the inland sites, but increases for the sites near the coast.

\begin{figure}[t!]
{\centering\includegraphics[scale=0.77]{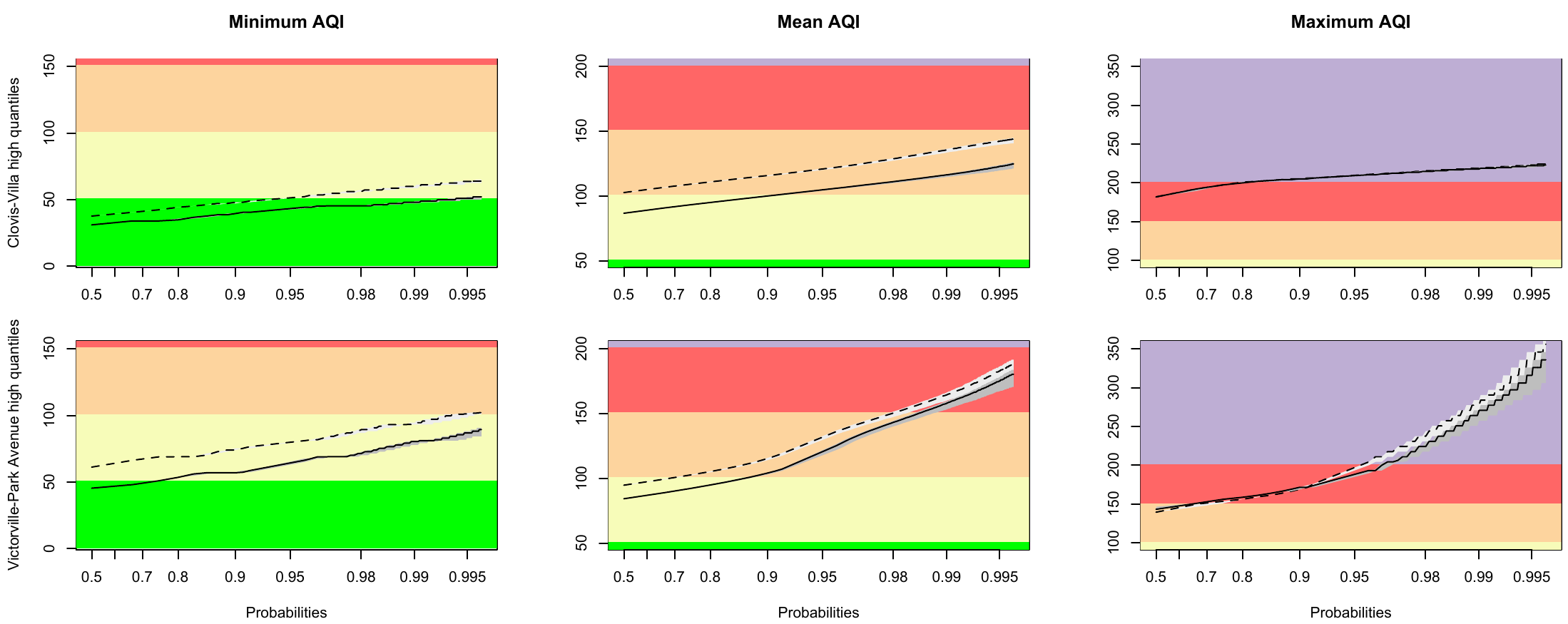}}
\caption{\footnotesize{High $p$-quantiles $z_p$ computed for the minimum (left), the average (center) and the maximum (right) AQI setting August 2006 (dashed lines) or August 2014 (solid lines) as baselines with 95\% bootstrap credible bands (grey areas) for CO, O$_3$ and NO$_2$ indexes, obtained for the Clovis-Villa (top) and Victorville-Park Avenue (bottom) site applying the TM-MCMC algorithm after $R=50000$ iterations and burn-in $R/5$ iterations. AQI categories: 0-50 satisfactory (green);  51-100 acceptable (yellow); 101-150 unhealthy for sensitive groups (orange); 151-200 unhealthy (red); $>$200 very unhealthy (purple). Probabilities are displayed on a Gumbel scale, i.e., $z_p$ is plotted against $-\log\{-\log(p)\}$.
}}\label{fig: returnlevels}
\end{figure}

The Air Quality Index (AQI) is a measure typically used by the U.S. EPA to communicate the level of air pollution to the public. When multiple criteria pollutants are measured at the same location, the EPA reports the largest AQI value as a measure of air quality; for more details see, e.g., \citet{AQIcate}. The AQI is divided into different categories that indicate different levels of health concern. Using our algorithm, we are able to sample from the posterior predictive distribution of the AQI values for the maxima of the criteria pollutants CO, O$_3$ and NO$_2$, taking into account the uncertainty associated with the trees summarising the dependence relations between these pollutants, meteorological parameters and other air pollutants using model averaging. 
In Figure~\ref{fig: returnlevels} we display high $p$-quantiles with probabilities ranging from $p=0.5$ to $p=0.996$, considering August 2006 and August 2014 as baselines, computed for the smallest, the average and the largest AQI monthly maxima for CO, O$_3$ and NO$_2$. The projections and credible bands for the AQI of CO, O$_3$ and NO$_2$ were computed as explained in \citet{epa2016} on the basis of 500 new datasets simulated from the nested logistic distribution with the dependence parameters obtained by 500 independent TM-MCMC algorithm runs with $R=50000$ and burn-in $R/5$. Notice that the values of $p=1-1/12\approx0.917$ and $p=1-1/(12\times 20)\approx 0.996$ roughly correspond to $1$ and $20$ year-return levels, respectively, under stationary conditions. The AQI categories are represented by different colours. For comparison purposes, we also computed high quantiles for the site named Victorville-Park Avenue. Since we estimate the dependence structure separately from the margins, the 95\% bootstrap credible intervals solely reflect the model uncertainty associated with the clustering, illustrating the ability of the TM-MCMC algorithm to realistically account for the model uncertainty in predictions. In practice, it might be better to adopt a fully Bayesian approach aggregating the margins and dependence structure uncertainties. Interestingly, the 95\% bootstrap credible bands are very narrow in the case of the Clovis-Villa site. Indeed, from Figure~\ref{fig: TREE}, a single tree structure (tree B) is sufficient to represent the whole dependence structure. 
On the other hand, as shown in Figure \ref{fig: TREE}, the site Victorville-Park Avenue is characterised by a higher model uncertainty, which reflects on much wider credible bands for the high quantiles in Figure~\ref{fig: returnlevels}.  
The high $p$-quantiles projections calculated based on August 2006 are generally larger than the high quantiles based on August 2014.  For 2014, the minimum AQI exceeds the satisfactory level of 50 with probability about $0.5\%$ at Clovis-Villa and around $20\%$ at Victorville-Park Avenue, i.e., once every $15$-$20$ years and $3$ times per year on average, respectively, under stationarity. Moreover, the average AQI generally lies within the unhealthy category only for sensitive groups of people in Clovis-Villa whereas at Victorville-Park Avenue it is expected to exceed this category with probability about $1.5\%$, so roughly once every 5 years on average. The maximum AQI high quantiles generally lie within the very unhealthy category, indicating that at least one of the criteria pollutants under study exceeds this most critical threshold with probability about $15\%$ at Clovis-Villa and around $2.5\%$ at Victorville-Park Avenue, and therefore $2$ times per year and once every 3 or 4 years on average, respectively. 

\subsection{Analysis of Clovis-Villa air pollution extremes}
\begin{figure}[t!]
\hspace{1.2cm}{\centering\includegraphics[scale=0.55]{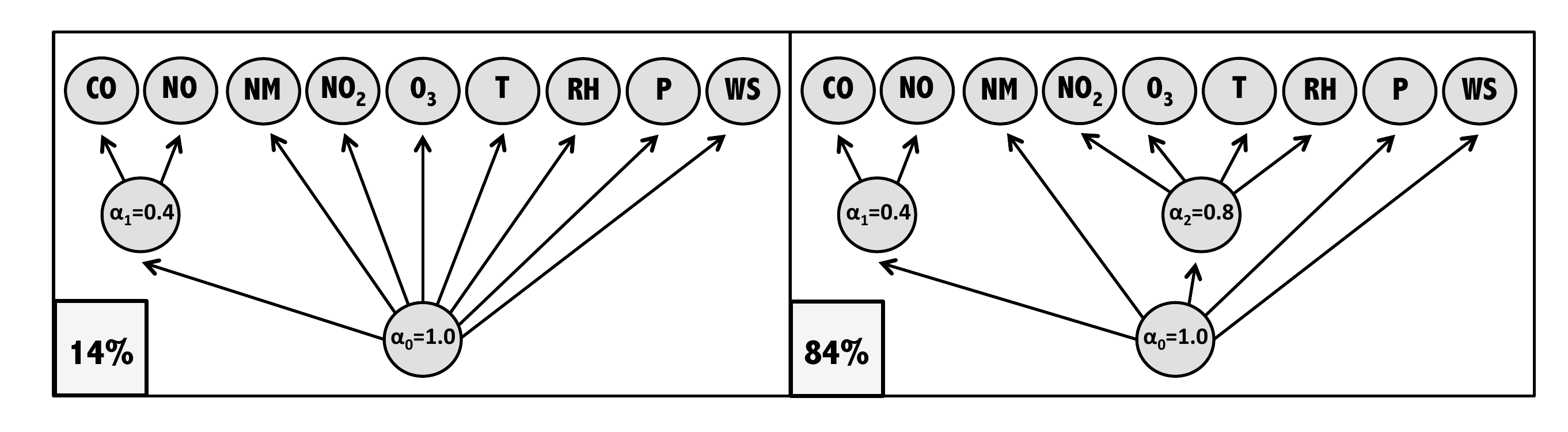}}
\caption{\footnotesize{Dependence structures identified by the TM-MCMC algorithm after $R=15000$ iterations and the associated posterior probability for Clovis-Villa.}}\label{fig: CVtree}
\end{figure}

We now include in our analysis the monthly maxima of relative humidity (RH), barometric pressure (BP) and wind speed (WS), together with concentration maxima of non-methane organic compounds (NM), which are essentially VOCs without methane. For illustrative purposes, we focus on the site named Clovis-Villa, located in Fresno, for which we were able to derive 54 multivariate monthly maxima, collected from January 2006 to December 2014.  The two tree structures identified by the TM-MCMC algorithm are illustrated in Figure~\ref{fig: CVtree}.
In accordance with previous findings, our algorithm groups the maxima of NO and CO through 98\% of the chain, and the maxima of O$_3$, NO$_2$ and T, now combined with RH, through 84\% of the chain. 
\begin{figure}[t!]
{\centering\includegraphics[scale=0.55]{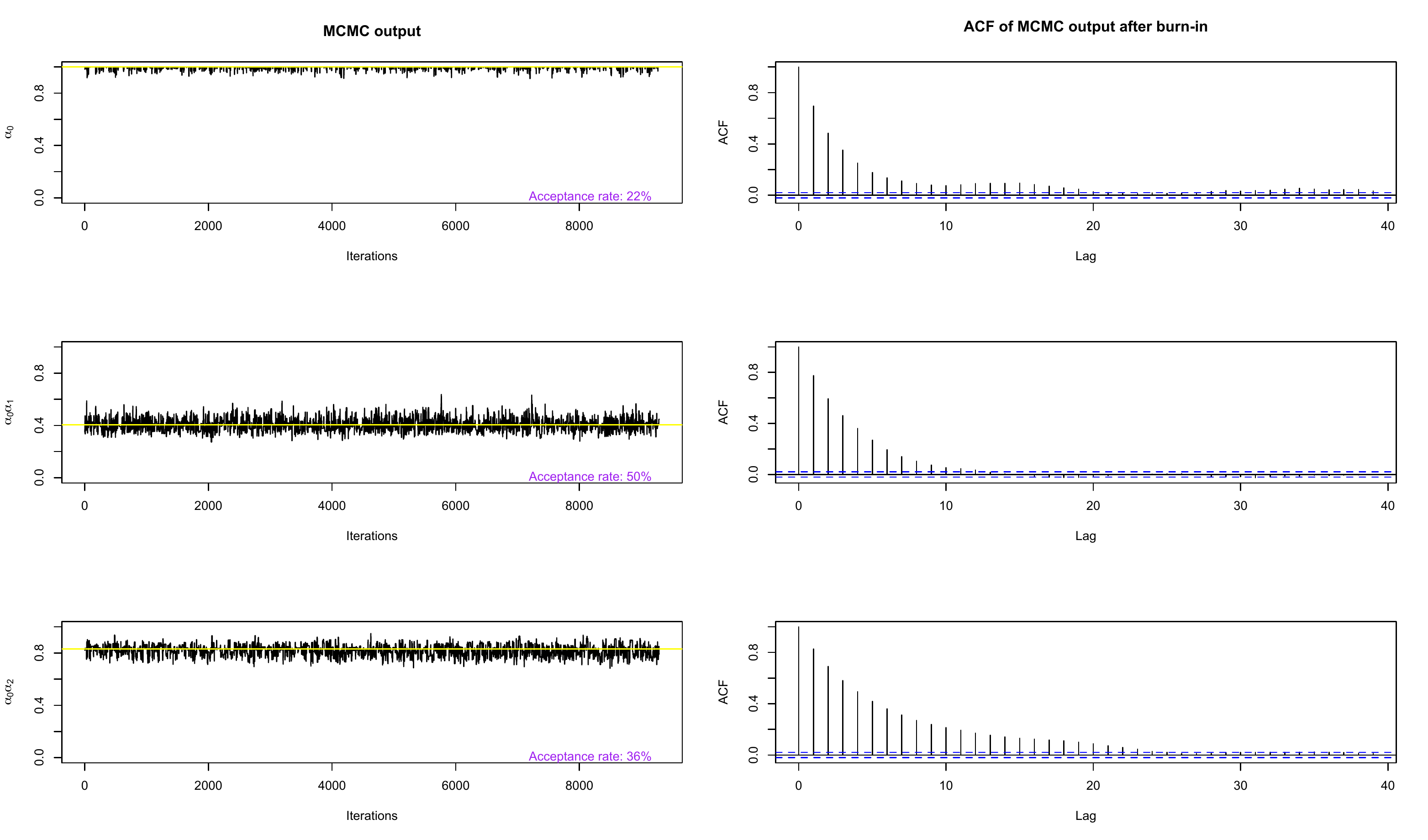}}
\caption{\footnotesize{Trace-plots (left) and the autocorrelation functions (right) for the sub-chain corresponding to the tree structure represented in Figure \ref{fig: CVtree}, with posterior probability of 84\% obtained by applying the TM-MCMC algorithm after $R=15000$ iterations after a burn-in of $R/5$ iterations. The posterior medians are represented by yellow lines. }}\label{fig: CVoutput}
\end{figure}
\begin{figure}[t!]
{\centering\includegraphics[scale=0.53]{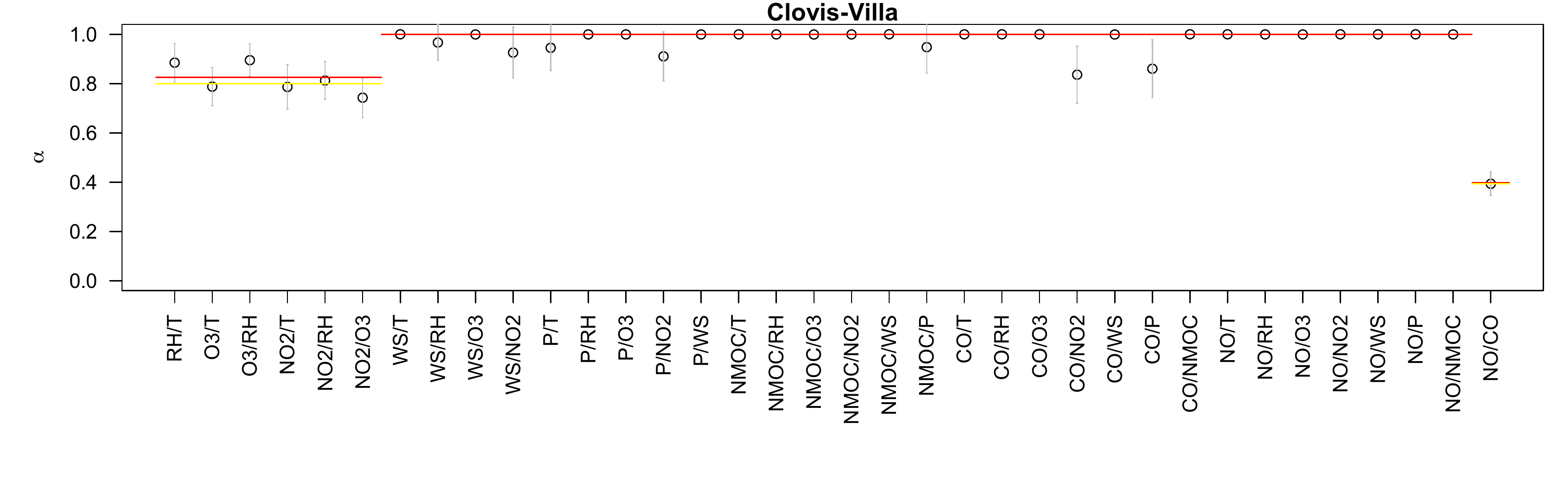}}
\caption{\footnotesize{Parameters estimates (circles) and 95\% credible intervals (vertical segments) resulting from logistic model bivariate fits for each pair of pollutants and meteorological parameters. The red lines represent the median of the parameter estimates and the yellow lines the posterior medians obtained from the multivariate fit using the TM-MCMC algorithm assuming the tree in the right panel of Figure \ref{fig: CVtree}, after $R=15000$ iterations and burn-in $R/5$ iterations. 
}}\label{fig: CVestimates}
\end{figure}

The algorithm output is provided in Figure~\ref{fig: CVoutput}. The sub-chains seem stationary from the trace-plots on the left panels and the autocorrelation functions on the right panels hint towards good sub-chain's mixing properties. In particular, the posterior median of the parameter $\alpha_0$ estimated by the algorithm is exactly 1, indicating that the two clusters of variables, as well as the other single variables, can be treated independently. Therefore, the algorithm allows us to significantly reduce the complexity of the data describing their extremal dependence structure by using only three parameters, or even two if $\alpha_0=1$, for the nine variables considered here. In contrast, fitting the bivariate logistic model to all possible pairs of variables yields 36 estimated dependence parameters, as shown in Figure~\ref{fig: CVestimates}. 
The bivariate fits suggest a relation of strong dependence between the maxima of NO and CO and of moderate dependence between the maxima of O$_3$, NO$_2$, T and RH, as indicated by the point estimates obtained from the algorithm output. Interestingly, the joint fits match the bivariate fits almost perfectly, except for a few pairs of variables with high variability.

\begin{figure}[t!]
{\centering\includegraphics[scale=0.485]{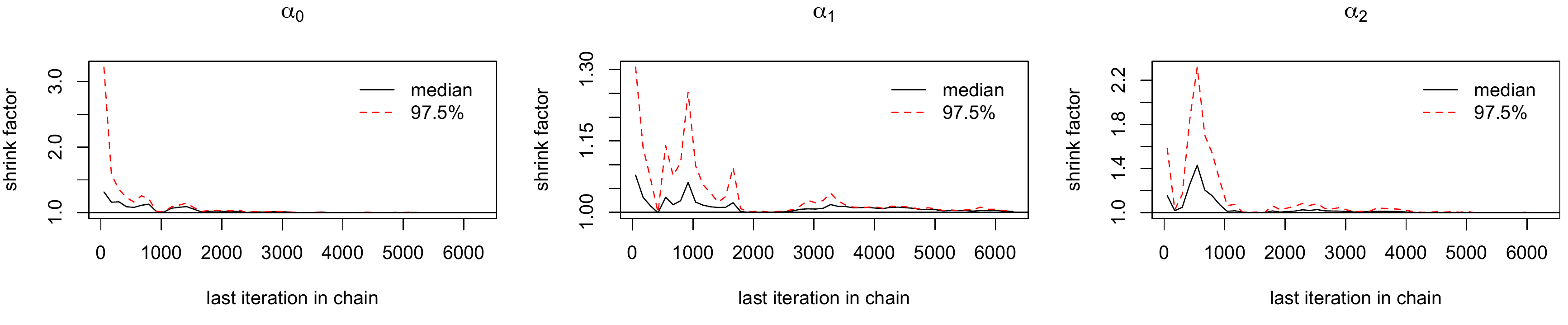}}
\caption{\footnotesize{Gelman-Rubin statistics \citep{Gelman1992} comparing two parallel sub-chains obtained from two independent runs of the TM-MCMC algorithm for $R=15000$ iterations and burn-in $R/5$.
}}\label{fig: diag}
\end{figure}

The Gelman--Rubin statistics \citep{Gelman1992}, plotted in Figure~\ref{fig: diag}, provide a convergence diagnostic measure based on the fact that if two different runs of the same model have converged, we expect the respective sub-chains to be similar to one another and the Gelman--Rubin statistics should tend to one as we increase the number of iterations. Since this is the case in Figure~\ref{fig: diag}, we conclude that there is no significant difference between the variance within and between the two algorithm sub-chains for each of the dependence parameters.
\section{Discussion} \label{sec: discussion}

In order to describe the complex dependence relationships between extremes variables, we proposed a novel technique based on the hierarchical structure of the nested logistic model and Bayesian computational tools. Our algorithm has the advantage of allowing parameter estimation and model selection to be done simultaneously, thus providing a Bayesian assessment of the model uncertainty. Numerical experiments suggest that, most of the time, the algorithm is able to find the true dependence structure from the data, even in case of model misspecification. 

The TM-MCMC algorithm was applied to air pollution concentration maxima collected at 21 sites across California.  
A particularly strong dependence relation between the maxima of CO and NO was found for most of the sites located between or within big cities, possibly because both pollutants are released by motor vehicles. Moreover, many of the sites close to coastal areas present a strong to mild dependence relation between the maxima of O$_3$ and NO$_2$ and high temperatures. This is expected as O$_3$ generally forms in chemical reactions that depend on the presence of NO$_2$ and heat. Fitting the bivariate logistic model to all possible pairs of variables for the Clovis-Villa site in Fresno yields similar conclusions, verifying that, despite its simplicity, the nested logistic model was flexible enough to provide a good estimation of the overall data dependence structure. 
The AQI high quantiles indicate that air quality is generally expected to exceed the healthy threshold within the next two decades under stationarity. Further efforts to reduce emissions seem necessary in order to protect public health, especially given that longer-term climatic changes projections predict rising temperatures. 
Our method could be used for obtaining air quality measures that take into account the extremes of multiple pollutants and the public health risks associated with their exposure. 

When fitting the nested logistic model or the nested Gumbel copula, its within-cluster exchangeability may be seen as a limitation. However, embedding the nested logistic distribution within a Bayesian framework has the great advantage of describing the data dependence structure using a mixture of trees, allowing therefore to capture complex dependence relationships between variables. Indeed, our method is not only able to appropriately describe data dependence structures that can be represented by homogeneous clusters, e.g., at the Clovis-Villa site, but also much more complicated dependencies, which often arise in applications, such as the situation at the Victorville-Park Avenue. 

Our method is coded in {\tt C++}, integrated within {\tt R} using the package {\tt Rcpp}. The code is available in the Supplementary Material. The computational efficiency remains fairly moderate and limited to a few seconds in the settings we have considered. In order to further improve the computational efficiency in high dimensions and facilitate convergence, a guided procedure could be implemented for selecting the move type at each iteration of the algorithm and additional moves might be implemented. 

A similar procedure might be applied to threshold exceedances based on the original series of observations, instead of on maxima, by implementing a censored recursive likelihood function, which may lead to further computational improvements. 

Our method is based on the nested logistic model, whose hierarchical structure is less appropriate to describe spatial extremes. In \cite{Vettori2018}, the nested logistic model is extended to the multivariate spatial framework, allowing for much large dimensions $D$.

Because of its max-stability property, the nested logistic model assumes asymptotic dependence between the margins, which might lead to overestimation of joint-tail probabilities under asymptotic independence \citep{Davison2013}. Multivariate models for asymptotic independence were proposed among others by  \citet*{Ledford1996,Ledford1997} and \citet{Hefferman2004}; see  \citet{Carvalho2012b} for a review. More recently, there has been an increasing research effort in bridging asymptotic dependence and independence in spatial \citep{Huser2017a,Huser2017b,Huser2018} and multivariate \citep{Wadsworth2017} settings, although the latter mostly focus on the bivariate case.
When it is not clear whether extreme data are asymptotically dependent or independent, \citet{Davison2013} suggest using max-stable models since the latter provide more conservative bounds for probabilities of concurrent extremes. In this paper, we analysed multivariate, a priori unstructured, maxima data, and we therefore chose to work with max-stable models. It would be interesting to generalise our algorithm to handle asymptotic independence scenarios.

\baselineskip=5pt
\bibliography{Bibliography}
\bibliographystyle{CUP}

\begin{appendix} \label{sec:appendices}
\baselineskip=18pt
\normalsize
\section{Recursive formulas for the nested logistic distribution}\label{AppendixA} 

\subsection{Notation}\label{sec:def}
For simplicity, we denote the distribution of the nested logistic model as
$G=\exp(-V),$ where 
$$V=\left(\sum_{k=1}^KV_k\right)^{\alpha_0},\qquad V_k=\left\{\sum_{i_k=1}^{D_k}z_{k;i_k}^{-1/(\alpha_0\alpha_k)}\right\}^{\alpha_k},\qquad k=1,2,\ldots,K,$$
with dependence parameters $0<\alpha_0,\alpha_1,\ldots,\alpha_K\leq1$, and where, for clarity, we have omitted the function arguments. Moreover, we use the following vector notation:
\vspace{-0.2cm}
\begin{align*}
\boldsymbol{z}_k&=(z_{k;1},\ldots,z_{k;D_k})^\top&&\mbox{\emph{All variables in cluster $k=1,\ldots,K$;}}\\
\boldsymbol{z}_{k,{1:d_k}}&=(z_{k;1},\ldots,z_{k;d_k})^\top&&\mbox{\emph{sub-vector of the first $d_k$ variables in cluster $k=1,\ldots,K$;}}\\
\boldsymbol{i}_{1:\kappa}&=(i_1,\ldots,i_\kappa)^\top&&\mbox{\emph{Vector of $\kappa$ indices.}}\\
\end{align*}
\vspace{-2cm}
\subsection{Preliminary results}
From the definition in Section \ref{sec:def}, we deduce the following derivatives:
\begin{eqnarray}
{\partial V_k\over\partial z_{k;i_k}}&=&\alpha_k\left\{\sum_{i_k=1}^{D_k}z_{k;i_k}^{-1/(\alpha_0\alpha_k)}\right\}^{\alpha_k-1}\times {-1\over\alpha_0\alpha_k}z_{k;i_k}^{-1/(\alpha_0\alpha_k)-1}\;\;=\;\;-{1\over\alpha_0} z_{k;i_k}^{-1/(\alpha_0\alpha_k)-1} V_k^{1-1/\alpha_k}\label{dVk};\\
{\partial V\over\partial z_{k;i_k}}&=&\alpha_0 \left(\sum_{k=1}^KV_k\right)^{\alpha_0-1}{\partial V_k\over\partial z_{k;i_k}}\;\;=\;\;-z_{k;i_k}^{-1/(\alpha_0\alpha_k)-1}V_k^{1-1/\alpha_k}V^{1-1/\alpha_0}\label{dV};\\
{\partial G\over\partial z_{k;i_k}}&=&-{\partial V\over\partial z_{k;i_k}}\exp(-V)\;\;=\;\;z_{k;i_k}^{-1/(\alpha_0\alpha_k)-1}G V_k^{1-1/\alpha_k}V^{1-1/\alpha_0}\label{dG}.
\end{eqnarray}

\subsection{Partial derivatives of G}\label{sec:proof}
The distribution $G$ is a function of $D=\sum_{k=1}^K D_k$ variables, namely $\boldsymbol{z}_1=(z_{1;1},\ldots,z_{1;D_1})^\top,\ldots,\boldsymbol{z}_K=(z_{K;1},\ldots,z_{K;D_K})^\top$. The partial derivative of $G$ with respect to any subset of variables $\boldsymbol{z}_{1,{1:d_1}},\ldots,\boldsymbol{z}_{\kappa,{1:d_\kappa}}$ in $1\leq\kappa\leq K$ clusters of dimensions $1\leq d_k\leq D_k$, $k=1,\ldots,\kappa$, may be expressed as
\begin{equation}\label{eq: recurs_partial_lik}
\begin{aligned}
\frac{\partial^{\sum_{k=1}^\kappa d_k}G}{\partial \prod_{k=1}^\kappa \mathbf{z}_{k;1:d_k}}=G\prod_{i_1=1}^{d_1}z_{1;i_1}^{-\frac{1}{\alpha_0\alpha_1}-1}\cdots\prod_{i_\kappa=1}^{d_\kappa}z_{\kappa;i_\kappa}^{-\frac{1}{\alpha_0\alpha_\kappa}-1} \sum_{i_{1}=1}^{d_{1}}\cdots\sum_{i_{\kappa}=1}^{d_{\kappa}}\sum_{j=1}^{\sum_{k=1}^\kappa i_k}\beta_{i_{1};\ldots;i_{\kappa};j}^{(d_{1};\ldots;d_{\kappa})}V_{1}^{i_{1}-\frac{d_{1}}{\alpha_1}}\cdots V_{\kappa}^{i_{\kappa}-\frac{d_{\kappa}}{\alpha_k}}V^{j-\frac{\sum_{k=1}^\kappa i_k}{\alpha_0}}
\end{aligned}
\end{equation}
where the coefficients $\beta_{i_{1};\ldots;i_{\kappa};j}^{(d_{1};\ldots;d_{\kappa})}$ can be computed recursively as demonstrated below.
By specialising the expression in \eqref{eq: recurs_partial_lik} to $\kappa=K$ clusters and $d_1=D_1,\ldots,d_K=D_K$ variables, we obtain the density, or full likelihood, for one replicate.

\paragraph{Proof}\label{sec:proof.1}
Equation \eqref{eq: recurs_partial_lik} may be proven by double induction over $\kappa\in\{1,\ldots,K\}$ and $d_k\in\{1,\ldots,D_k\}$, $k=1,\ldots,\kappa$. The proof also naturally provides a constructive approach to the recursive computation of the coefficients $\beta_{i_{1};\ldots;i_{\kappa};j}^{(d_{1};\ldots;d_{\kappa})}$. More precisely, we demonstrate the following four steps:
\begin{enumerate}
\item \eqref{eq: recurs_partial_lik} holds for $\kappa=1$ and $d_1=1$;
\item If \eqref{eq: recurs_partial_lik} holds for $\kappa=1$ and $d_1\in\{1,\ldots,D_1-1\}$, then it also holds for $d_1\mapsto d_1+1$;
\item If \eqref{eq: recurs_partial_lik} holds for $\kappa\in\{1,\ldots,K-1\}$, then it also holds for $\kappa\mapsto \kappa+1$ with $d_{\kappa+1}=1$;
\item If \eqref{eq: recurs_partial_lik} holds for $\kappa\in\{1,\ldots,K\}$ and $d_\kappa\in\{1,\ldots,D_{\kappa-1}\}$, then it also holds for $d_\kappa\mapsto d_{\kappa}+1$.
\end{enumerate}
\paragraph{Step 1.}
From \eqref{dG} we have
$${\partial G\over\partial z_{1;1}}=z_{1;1}^{-1/(\alpha_0\alpha_1)-1} G V_1^{1-1/\alpha_1}V^{1-1/\alpha_0},$$
which proves the first step by setting $\beta_{1;1}^{(1)}=1$.

\paragraph{Step 2.}\label{step2}
Assuming that \eqref{eq: recurs_partial_lik} holds for $\kappa=1$ and $d_1\in\{1,\ldots,D_1-1\}$, and using \eqref{dVk}, \eqref{dV} and \eqref{dG}, we obtain
\[
\begin{aligned}
&\frac{\partial^{d_{1}+1}G}{\partial \mathbf{z}_{1;1:d_1+1}}=\prod_{i_1=1}^{d_1}z_{1;i_1}^{-\frac{1}{\alpha_0\alpha_1}-1}\left\{\frac{\partial G}{z_{1;d_{1}+1}}\sum_{i_{1}=1}^{d_{1}}\sum_{j=1}^{i_{1}}\beta_{i_{1};j}^{(d_{1})}V_{1}^{i_{1}-\frac{d_{1}}{\alpha_1}}V^{j-\frac{i_{1}}{\alpha_0}}\right.
+G\sum_{i_{1}=1}^{d_{1}}\sum_{j=1}^{i_{1}}\beta_{i_{1};j}^{(d_{1})}\frac{\partial V_{1}^{i_{1}-\frac{d_{1}}{\alpha_1}}}{z_{1;d_{1}+1}}V^{j-\frac{i_{1}}{\alpha_0}}\\
&\left.+G\sum_{i_{1}=1}^{d_{1}}\sum_{j=1}^{i_{1}}\beta_{i_{1};j}^{(d_{1})}V_{1}^{i_{1}-\frac{d_{1}}{\alpha_1}}\frac{\partial V^{j-\frac{i_{1}}{\alpha_0}}}{z_{1;d_{1}+1}}\right\}
 \end{aligned}
\]
\[
\begin{aligned}
=&\;G\prod_{i_1=1}^{d_1+1}z_{1;i_1}^{-\frac{1}{\alpha_0\alpha_1}-1}\left\{\sum_{i_{1}=1}^{d_{1}}\sum_{j=1}^{i_{1}}\beta_{i_{1};j}^{(d_{1})}V_{1}^{\left(i_{1}+1\right)-\frac{\left(d_{1}+1\right)}{\alpha_1}}V^{\left(l+1\right)-\frac{\left(i_{1}+1\right)}{\alpha_0}}\right.\\
&\left.-\sum_{i_{1}=1}^{d_{1}}\sum_{j=1}^{i_{1}}\beta_{i_{1};j}^{(d_{1})}\left(i_{1}-\frac{d_{1}}{\alpha_1}\right)\left(\frac{1}{\alpha_0}\right)V_{1}^{i_{1}-\frac{d_{1}+1}{\alpha_1}}V^{j-\frac{i_{1}}{\alpha_0}}-\sum_{i_{1}=1}^{d_{1}}\sum_{j=1}^{i_{1}}\beta_{i_{1};j}^{(d_{1})}\left(l-\frac{i_{1}}{\alpha_0}\right)V_{1}^{\left(i_{1}+1\right)-\frac{d_{1}+1}{\alpha_1}}V^{j-\frac{\left(i_{1}+1\right)}{\alpha_0}}\right\}\\
=&\;G\prod_{i_1=1}^{d_1+1}z_{1;i_1}^{-\frac{1}{\alpha_0\alpha_1}-1}\left\{\sum_{i_{1}=2}^{d_{1}+1}\sum_{l=2}^{i_{1}}\beta_{i_{1}-1;j-1}^{(d_{1})}V_{1}^{i_{1}-\frac{\left(d_{1}+1\right)}{\alpha_1}}V^{j-\frac{i_{1}}{\alpha_0}}\right.-\sum_{i_{1}=1}^{d_{1}}\sum_{j=1}^{i_{1}}\beta_{i_{1};j}^{(d_{1})}\left(i_{1}-\frac{d_{1}}{\alpha_1}\right)\left(\frac{1}{\alpha_0}\right)V_{1}^{i_{1}-\frac{d_{1}+1}{\alpha_1}}V^{j-\frac{i_{1}}{\alpha_0}}\\
&\left.-\sum_{i_{1}=2}^{d_{1}+1}\sum_{j=1}^{i_{1}}\beta_{i_{1}-1;j}^{(d_{1})}\left(l-\frac{i_{1}-1}{\alpha_0}\right)V_{1}^{i_{1}-\frac{d_{1}+1}{\alpha_1}}V^{j-\frac{i_{1}}{\alpha_0}}\right\}
=G\prod_{i_1=1}^{d_1+1}z_{1;i_1}^{-\frac{1}{\alpha_0\alpha_1}-1} \sum_{i_{1}=1}^{d_{1}+1}\sum_{j=1}^{i_{1}}\beta_{i_{1};j}^{(d_{1}+1)}V_{1}^{i_{1}-\frac{d_{1}+1}{\alpha_1}}V^{j-\frac{i_{1}}{\alpha_0}},
 \end{aligned}
\]
where
\begin{equation}\label{recurs_beta1}
\beta_{i_{1};j}^{(d_{1}+1)}=
\beta_{i_{1}-1;j-1}^{(d_{1})} -\frac{1}{\alpha_0}\left(i_{1}-\frac{d_{1}}{\alpha_1}\right)\beta_{i_{1};j}^{(d_{1})} -\left(l-\frac{i_{1}-1}{\alpha_0}\right)\beta_{i_{1}-1;j}^{(d_{1})},  \quad 1\leq j\leq  i_{1}\leq d_{1}+1,
\end{equation}
with
\begin{equation}\label{recurs_beta2}
\beta_{i_1;j}^{(d_1)}=0,\qquad\mbox{for all}\quad i_1\notin\{1,\ldots,d_1\},\quad \mbox{or}\quad j\notin\{1,\ldots,i_1\}.
\end{equation}
Hence, \eqref{eq: recurs_partial_lik} holds by induction for $\kappa=1$ and  any $1\leq d_1\leq D_1$, and the recursive formula to compute coefficients $\beta_{i_1;j}^{(d_1)}$ is given by \eqref{recurs_beta1} and \eqref{recurs_beta2} with the initial condition $\beta_{1;1}^{(1)}=1$. 

\paragraph{Step 3.}
Assuming that \eqref{eq: recurs_partial_lik} holds for $\kappa\in\{1,\ldots,K-1\}$, we obtain
\[
\begin{aligned}
&\frac{\partial^{1+\sum_{k=1}^\kappa d_{k}}G}{\partial \prod_{k=1}^\kappa \mathbf{z}_{k;1:d_k}\partial z_{\kappa+1;1}}=\prod_{i_1=1}^{d_1}z_{1;i_1}^{-\frac{1}{\alpha_0\alpha_1}-1}\cdots\prod_{i_\kappa=1}^{d_\kappa}z_{\kappa;i_\kappa}^{-\frac{1}{\alpha_0\alpha_\kappa}-1} \left\{\frac{\partial G}{\partial z_{\kappa+1;1}}\sum_{i_{1}=1}^{d_{1}}\cdots\sum_{i_{\kappa}=1}^{d_{\kappa}}V_{1}^{i_{1}-\frac{d_{1}}{\alpha_1}}\cdots V_{k}^{i_{\kappa}-\frac{d_{\kappa}}{\alpha_k}}\right.\\
&\left.\times\sum_{j=1}^{i_{1}+\cdots+i_{\kappa}}\beta_{i_{1};\ldots;i_{\kappa};j}^{(d_{1};\ldots;d_{\kappa})}V^{j-\frac{i_{1}+\cdots+i_{\kappa}}{\alpha_0}}+G\sum_{i_{1}=1}^{d_{1}}\cdots\sum_{i_{\kappa}=1}^{d_{\kappa}}\sum_{j=1}^{i_{1}+\cdots+i_{\kappa}}\beta_{i_{1};\ldots;i_{\kappa};j}^{(d_{1};\ldots;d_{\kappa})}V_{1}^{i_{1}-\frac{d_{1}}{\alpha_1}}\cdots V_{k}^{i_{\kappa}-\frac{d_{\kappa}}{\alpha_k}}\frac{\partial V^{j-\frac{i_{1}+\cdots+i_{\kappa}}{\alpha_0}}}{\partial z_{\kappa+1;1}}\right\}\\
=&\;G\prod_{i_1=1}^{d_1}z_{1;i_1}^{-\frac{1}{\alpha_0\alpha_1}-1}\cdots\prod_{i_\kappa=1}^{d_\kappa}z_{\kappa;i_\kappa}^{-\frac{1}{\alpha_0\alpha_\kappa}-1} (z_{\kappa+1;1})^{-\frac{1}{\alpha_0\alpha_{\kappa+1}}-1} \sum_{i_{1}=1}^{d_{1}}\cdots\sum_{i_{\kappa}=1}^{d_{\kappa}}V_{1}^{i_{1}-\frac{d_{1}}{\alpha_1}}\cdots V_{k}^{i_{\kappa}-\frac{d_{\kappa}}{\alpha_k}}\left\{\sum_{l=2}^{i_{1}+\cdots+i_{\kappa}+1}\beta_{i_{1};\ldots;i_{\kappa};j-1}^{(d_{1};\ldots;d_{\kappa})}\right.\\
&\times\left.V_{\kappa+1}^{1-\frac{1}{\alpha_{\kappa+1}}}V^{j-\frac{i_{1}+\cdots+i_{\kappa}+1}{\alpha_0}}- \sum_{j=1}^{i_{1}+\cdots+i_{\kappa}}\beta_{i_{1};\ldots;i_{\kappa};j}^{(d_{1};\ldots;d_{\kappa})}\left(l-\frac{i_{1}+\cdots+i_{\kappa}}{\alpha_0}\right)V_{\kappa+1}^{1-\frac{1}{\alpha_{\kappa+1}}}V^{j-\frac{i_{1}+\cdots+i_{\kappa}+1}{\alpha_0}}\right\}\\
=&\;G\prod_{i_1=1}^{d_1}z_{1;i_1}^{-\frac{1}{\alpha_0\alpha_1}-1}\cdots\prod_{i_\kappa=1}^{d_\kappa}z_{\kappa;i_\kappa}^{-\frac{1}{\alpha_0\alpha_\kappa}-1} (z_{\kappa+1;1})^{-\frac{1}{\alpha_0\alpha_{\kappa+1}}-1} \\
&\times\sum_{i_{1}=1}^{d_{1}}\cdots\sum_{i_{\kappa}=1}^{d_{\kappa}}\sum_{j=1}^{i_{1}+\cdots+i_{\kappa}+1}\beta_{i_{1};\ldots;i_{\kappa};1;j}^{(d_{1};\ldots;d_{\kappa};1)}V_{1}^{i_{1}-\frac{d_{1}}{\alpha_1}}\cdots V_{k}^{i_{\kappa}-\frac{d_{\kappa}}{\alpha_k}}V_{\kappa+1}^{1-\frac{1}{\alpha_{\kappa+1}}}V^{j-\frac{i_{1}+\cdots+i_{\kappa}+1}{\alpha_0}},
\end{aligned}
\]
where
\begin{equation}\label{eq: recurs_beta3a}
\beta_{i_{1};\ldots;i_{\kappa};1;j}^{(d_{1};\ldots;d_{\kappa};1)}=
\beta_{i_{1};\ldots;i_{\kappa};j-1}^{(d_{1};\ldots;d_{\kappa})}-\left(l-\frac{i_{1}+\cdots+i_{\kappa}}{\alpha_0}\right)\beta_{i_{1};\ldots;i_{\kappa};j}^{(d_{1};\ldots;d_{\kappa})}, 1\leq j\leq \sum^{\kappa}_{k=1}i_{k}+1, 1\leq i_{k}\leq d_{k}, \;k=1,\ldots,\kappa,
\end{equation}
with 
\begin{equation}\label{eq: recurs_beta3b}
\beta_{i_{1};\ldots;i_{\kappa};1;j}^{(d_{1};\ldots;d_{\kappa};1)}=0,\quad\mbox{for all}\quad i_k\notin\{1,\ldots,d_k\}, \;k=1,\ldots,\kappa,\quad \mbox{or}\quad l\notin\{1,\ldots,i_{1}+\cdots+i_{\kappa}\}.
\end{equation}

Hence, \eqref{eq: recurs_partial_lik} holds by induction for $\kappa\mapsto\kappa+1$ with $d_{\kappa+1}=1$ and the recursive formula to compute coefficients $\beta_{i_{1};\ldots;i_{\kappa},1,l}^{(d_{1};\ldots;d_{\kappa},1)}$ is given by \eqref{eq: recurs_beta3a} and \eqref{eq: recurs_beta3b}. 

\paragraph{Step 4.}
Assuming that \eqref{eq: recurs_partial_lik} holds for $\kappa\in\{1,\ldots,K\}$ and $d_\kappa\in\{1,\ldots,D_\kappa-1\}$, we obtain
\[
\begin{aligned}
&\frac{\partial^{1+\sum_{k=1}^{\kappa} d_{k}}G}{\partial \prod_{k=1}^\kappa \mathbf{z}_{k;1:d_k}\partial z_{\kappa;d_{\kappa}+1}}=
\prod_{i_1=1}^{d_1}z_{1;i_1}^{-\frac{1}{\alpha_0\alpha_1}-1}\cdots\prod_{i_\kappa=1}^{d_\kappa}z_{\kappa;i_\kappa}^{-\frac{1}{\alpha_0\alpha_\kappa}-1} \left\{\frac{\partial G}{\partial z_{\kappa;d_{\kappa}+1}}\sum_{i_{1}=1}^{d_{1}}\cdots\sum_{i_{\kappa}=1}^{d_{\kappa}}\sum_{j=1}^{i_{1}+\cdots+i_{\kappa}}\beta_{i_{1};\ldots;i_{\kappa};j}^{(d_{1};\ldots;d_{\kappa})}\right.\\
\end{aligned}
\]
\[
\begin{aligned}
&\times V_{1}^{i_{1}-\frac{d_{1}}{\alpha_1}}\cdots V_{\kappa}^{i_{\kappa}-\frac{d_{\kappa}}{\alpha_k}}V^{j-\frac{i_{1}+\cdots+i_{\kappa}}{\alpha_0}}+G\sum_{i_{1}=1}^{d_{1}}\cdots\sum_{i_{\kappa}=1}^{d_{\kappa}}\sum_{j=1}^{i_{1}+\cdots+i_{\kappa}}\beta_{i_{1};\ldots;i_{\kappa};j}^{(d_{1};\ldots;d_{\kappa})}V_{1}^{i_{1}-\frac{d_{1}}{\alpha_1}}\cdots \frac{\partial V_{\kappa}^{i_{\kappa}-\frac{d_{\kappa}}{\alpha_{\kappa}}}}{\partial z_{\kappa;d_{\kappa}+1}} V^{j-\frac{i_{1}+\cdots+i_{\kappa}}{\alpha_0}}\\
&+\left.G\sum_{i_{1}=1}^{d_{1}}\cdots\sum_{i_{\kappa}=1}^{d_{\kappa}}\sum_{j=1}^{i_{1}+\cdots+i_{\kappa}}\beta_{i_{1};\ldots;i_{\kappa};j}^{(d_{1};\ldots;d_{\kappa})}V_{1}^{i_{1}-\frac{d_{1}}{\alpha_1}}\cdots  V_{\kappa}^{i_{\kappa}-\frac{d_{\kappa}}{\alpha_{\kappa}}} \frac{\partial V^{j-\frac{i_{1}+\cdots+i_{\kappa}}{\alpha_0}}}{\partial z_{\kappa;d_{\kappa}+1}}\right\}\\
=&\;G\prod_{i_1=1}^{d_1}z_{1;i_1}^{-\frac{1}{\alpha_0\alpha_1}-1}\cdots\prod_{i_\kappa=1}^{d_\kappa+1}z_{\kappa;i_\kappa}^{-\frac{1}{\alpha_0\alpha_\kappa}-1}  \left\{\sum_{i_{1}=1}^{d_{1}}\cdots\sum_{i_{\kappa}=2}^{d_{\kappa}+1}\sum_{l=2}^{i_{1}+\cdots+i_{\kappa}}\beta_{i_{1};\ldots;i_{\kappa}-1;j-1}^{(d_{1};\ldots;d_{\kappa})}V_{1}^{i_{1}-\frac{d_{1}}{\alpha_1}}\cdots V_{\kappa}^{i_{\kappa}-\frac{d_{\kappa}+1}{\alpha_{\kappa}}}V^{j-\frac{i_{1}+\cdots+i_{\kappa}}{\alpha_0}}\right.\\
&-\sum_{i_{1}=1}^{d_{1}}\cdots\sum_{i_{\kappa}=1}^{d_{\kappa}}\sum_{j=1}^{i_{1}+\cdots+i_{\kappa}}\beta_{i_{1};\ldots;i_{\kappa};j}^{(d_{1};\ldots;d_{\kappa})}V_{1}^{i_{1}-\frac{d_{1}}{\alpha_1}}\cdots \left(i_{\kappa}-\frac{d_{\kappa}}{\alpha_{\kappa}}\right) \left(\frac{1}{\alpha_0}\right) V_{\kappa}^{i_{\kappa}-\frac{d_{\kappa}+1}{\alpha_{\kappa}}}V^{j-\frac{i_{1}+\cdots+i_{\kappa}}{\alpha_0}}\\
&-\sum_{i_{1}=1}^{d_{1}}\cdots\sum_{i_{\kappa}=2}^{d_{\kappa}+1}\sum_{j=1}^{i_{1}+\cdots+i_{\kappa}-1}\beta_{i_{1};\ldots;i_{\kappa}-1;j}^{(d_{1};\ldots;d_{\kappa})}V_{1}^{i_{1}-\frac{d_{1}}{\alpha_1}}\cdots\left. V_{\kappa}^{i_{\kappa}-\frac{d_{\kappa}+1}{\alpha_{\kappa}}}\left(l-\frac{i_{1}+\cdots+i_{\kappa}-1}{\alpha_0}\right)V^{j-\frac{i_{1}+\cdots+i_{\kappa}}{\alpha_0}}\right\} \\
=&\;G\prod_{i_1=1}^{d_1}z_{1;i_1}^{-\frac{1}{\alpha_0\alpha_1}-1}\cdots\prod_{i_\kappa=1}^{d_\kappa+1}z_{\kappa;i_\kappa}^{-\frac{1}{\alpha_0\alpha_\kappa}-1}  \sum_{i_{1}=1}^{d_{1}}\cdots\sum_{i_{\kappa}=1}^{d_{\kappa}+1}\sum_{j=1}^{i_{1}+\cdots+i_{\kappa}}\beta_{i_{1};\ldots;i_{\kappa};j}^{(d_{1};\ldots;d_{\kappa}+1)}V_{1}^{i_{1}-\frac{d_{1}}{\alpha_1}}\cdots V_{k}^{i_{\kappa}-\frac{d_{\kappa}+1}{\alpha_k}}V^{j-\frac{i_{1}+\cdots+i_{\kappa}}{\alpha_0}}
\end{aligned}
\]
where
\begin{equation}\label{recurs_beta4a}
\beta_{i_{1};\ldots;i_{\kappa};j}^{(d_{1};\ldots;d_{\kappa}+1)}=
\beta_{i_{1};\ldots;i_{\kappa}-1;j-1}^{(d_{1};\ldots;d_{\kappa})}-\frac{1}{\alpha_0}\left(i_{\kappa}-\frac{d_{\kappa}}{\alpha_{\kappa}}\right)\beta_{i_{1};\ldots;i_{\kappa};j}^{(d_{1};\ldots;d_{\kappa})}-\left(l-\frac{i_{1}+\cdots+i_{\kappa}-1}{\alpha_0}\right)\beta_{i_{1};\ldots;i_{\kappa}-1;j}^{(d_{1};\ldots;d_{\kappa})},
\end{equation}
if $1\leq j\leq 1+\sum^{\kappa}_{k=1}i_{k}, i_{k}\leq d_{k}, \;k=1,\ldots,\kappa,$  with
\begin{equation}\label{recurs_beta4b}
\beta_{i_{1};\ldots;i_{\kappa};j}^{(d_{1};\ldots;d_{\kappa}+1)}=0, \quad\mbox{for all}\quad i_k\notin\{1,\ldots,d_k\}, \;k=1,\ldots,\kappa,\quad \mbox{or}\quad j\notin\{1,\ldots,\sum^{\kappa}_{k=1}i_{k}\}.
\end{equation}

Hence, \eqref{eq: recurs_partial_lik} holds by induction for $\in\{1,\ldots,K\}$ with $d_\kappa\mapsto d_{\kappa}+1$, and the recursive formula to compute coefficients $\beta_{i_{1};\ldots;i_{\kappa};j}^{(d_{1};\ldots;d_{\kappa}+1)}$ is given by \eqref{recurs_beta4a} and \eqref{recurs_beta4b}. 

\subsection{Complexity} 
If $\kappa=1$, the number of coefficients $\beta_{i_1;j}^{(d_1)}$ to be computed recursively in \eqref{eq: recurs_partial_lik} is
\[
\begin{aligned}
& \sum_{h=1}^{d_1}\left(\sum_{i_1=1}^h\sum_{j=1}^{i_1}1\right)=\sum_{h=1}^{d_1}\frac{h(h+1)}{2}=\frac{1}{2}\left(\sum_{h=1}^{d_1}h^2+\sum_{h=1}^{d_1}h\right)=
\frac{d_1(d_1+1)(d_1+2)}{6},
\end{aligned}
\]
which implies that the complexity is $\mathcal{O}(d_1^3)$. For $1\leq\kappa\leq K$ clusters of size $d_1,\ldots,d_\kappa$, the number of coefficients $\beta_{i_{1};\ldots;i_{\kappa};j}^{(d_{1};\ldots;d_{\kappa})}$ to be computed in \eqref{eq: recurs_partial_lik} is
\[
\begin{aligned}
&\sum_{h=1}^{d_\kappa}\left(\sum_{i_1=1}^{d_1}\cdots\sum_{i_{\kappa-1}=1}^{d_{\kappa-1}}\sum_{i_{\kappa}=1}^{h}\sum_{j=1}^{i_1+\cdots+i_\kappa}1\right)=\sum_{h=1}^{d_\kappa}\sum_{i_\kappa=1}^{h}\sum_{i_1=1}^{d_1}\cdots\sum_{i_{\kappa-1}=1}^{d_{\kappa-1}}(i_1+\cdots+i_\kappa)\\
=&\;\sum_{h=1}^{d_\kappa}\sum_{i_\kappa=1}^{h}\left\{\frac{d_1(d_1+1)}{2}d_1\cdots d_{\kappa-1}+\cdots+d_1\cdots d_{\kappa-2}\frac{d_{\kappa-1}(d_{\kappa-1}+1)}{2}+d_1\cdots d_{\kappa-1}i_\kappa\right\}\\
=&\;\left\{\frac{d_1(d_1+1)}{2}d_2\cdots d_{\kappa-1}\frac{d_\kappa(d_\kappa+1)}{2}\right\}+\cdots+\left\{d_1\cdots d_{\kappa-2}\frac{d_{\kappa-1}(d_{\kappa-1}+1)}{2}\frac{d_\kappa(d_\kappa+1)}{2}\right\}\\
&+ \left\{d_1\cdots d_{\kappa-1}\frac{d_\kappa(d_\kappa+1)(d_\kappa+2)}{6}\right\},
\end{aligned}
\]
which implies that the total complexity for the computation of the coefficients $\beta_{i_{1};\ldots;i_{\kappa};j}^{(d_{1};\ldots;d_{\kappa})}$ in \eqref{eq: recurs_partial_lik} is
\[
\mathcal{O}\left(\sum^\kappa_{k=1}(d_1+\cdots+d_k)d_1\cdots d_{k-1}d_k^2\right).
\]
If the cluster size is the same for all clusters, i.e., $d_k=d_1$, for all $k=2,\ldots,\kappa$, then we have
$\mathcal{O}\left(\kappa\sum^\kappa_{k=1}d_1^{k+2}\right).$ 
 \end{appendix}
 \end{document}